\documentclass{siamltex}
\usepackage{euscript,amsmath,amssymb,amsfonts,graphicx,bm}
  \usepackage{paralist}
  \usepackage{epstopdf}
  \usepackage{graphics} 
 \usepackage[colorlinks=true]{hyperref}
 \hypersetup{urlcolor=blue, citecolor=red}
\usepackage{cite}

\newcommand{\x}{{\bf x}}

\renewcommand{\d}{{\rm d}}
\newcommand{\e}{{\rm e}}

\newcommand{\Rset}{{\mathbb R}}

\newcommand{\mc}{\mathcal}
\newcommand{\pd}{\partial}

\title{Nonlinear Langevin equations for wandering patterns in stochastic neural fields}

\author{Paul C. Bressloff\thanks{Department of Mathematics, University of Utah, Salt Lake City, UT 84112 USA ({\tt bressloff@math.utah.edu})}\and Zachary P. Kilpatrick\thanks{Department of Mathematics, University of Houston, Houston, Texas 77204, USA ({\tt zpkilpat@math.uh.edu})}}

\date{\today}

\begin{document}

\maketitle
\newcommand{\slugmaster}{%
\slugger{MMedia}{xxxx}{xx}{x}}

\begin{abstract}
We analyze the effects of additive, spatially extended noise on spatiotemporal patterns in continuum neural fields. Our main focus is how fluctuations impact patterns when they are weakly coupled to an external stimulus or another equivalent pattern. Showing the generality of our approach, we study both propagating fronts and stationary bumps. Using a separation of time scales, we represent the effects of noise in terms of a phase-shift of a pattern from its uniformly translating position at long time scales, and fluctuations in the pattern profile around its instantaneous position at short time scales. In the case of a stimulus-locked front, we show that the phase-shift satisfies a nonlinear Langevin equation (SDE) whose deterministic part has a unique stable fixed point. Using a linear-noise approximation, we thus establish that wandering of the front about the stimulus-locked state is given by an Ornstein-Uhlenbeck (OU) process. Analogous results hold for the relative phase-shift between a pair of mutually coupled fronts, provided that the coupling is excitatory. On the other hand, if the mutual coupling is given by a Mexican hat function (difference of exponentials), the linear-noise approximation breaks down due to the co-existence of stable and unstable phase-locked states in the deterministic limit. Similarly, the stochastic motion of mutually coupled bumps can be described by a system of nonlinearly coupled SDEs, which can be linearized to yield a multivariate OU process. As in the case of fronts, large deviations can cause bumps to temporarily decouple, leading to a phase-slip in the bump positions.

\end{abstract}

\section{Introduction}

Several previous studies have explored the impact of fluctuations on spatiotemporal patterns in neural field equations by including a perturbative spatially extended noise term  \cite{Laing01,Hutt08,Bressloff12,Kilpatrick13}. Utilizing small-noise expansions, it is then possible to develop effective equations to describe the stochastic motion of spatiotemporal patterns that emerge in the noise-free system \cite{Sancho07}. Typically, these effective equations are linear stochastic differential equations (SDEs) \cite{Bressloff12,Kilpatrick13a}, although there have been derivations of stochastic amplitude equations in the vicinity of bifurcations \cite{Hutt08,Kilpatrick14} or nonlinear SDEs in spatially heterogeneous networks \cite{Kilpatrick13}. Here we demonstrate methods for deriving nonlinear SDEs for the effective motion of patterns in stochastic neural field equations, focusing on fronts and bumps in particular.

The basic neural field equation with a noise term $R(x,t)$ takes the form
\begin{eqnarray}
  \tau \frac{\partial u(x,t)}{\partial t}&=&-u(x,t)+\int_{\Omega} w(x-y)F(u(y,t))dydt+\sqrt{\epsilon}R(x,t),
\end{eqnarray}
where $u(x,t)$ represents neural population activity at position $x \in \Omega$ and time $t$, where $\Omega$ is a one-dimensional domain such as ${\mathbb R}$ or the ring $[- \pi, \pi]$. The function $F$ is a nonlinear firing-rate function. Synaptic connectivity in the network is represented by the function $w(x-y)$, which describes the polarity (sign) and strength (amplitude) of connection from location $y$ to $x$. The stochastic forcing is assumed to be weak by taking $0<\epsilon \ll 1$. Expressing $R(x,t)$ in terms of a spatially extended Wiener process, one can formulate the neural field equation as a stochastic integro-differential equation on a suitable function space such as ${\mathbb L}^2(\Rset)$. Suppose that the deterministic neural field equation ($\epsilon=0$) supports a wave solution $U_0(\xi)$ with $\xi=x-ct$ and $c$ the wave speed ($c \equiv 0$ for stationary waves such as bumps \cite{Kilpatrick13}). Motivated by perturbation methods for analyzing fronts in stochastic reaction-diffusion equations \cite{Mikhailov83,Sancho98,Sancho07}, we used a separation of time-scales to decompose the effects of noise into (i) a slow, diffusive-like displacement $\Delta(t)$ of the wave from its uniformly translating position, and (ii) fast fluctuations in the wave profile. More explicitly,
\begin{equation}
u(x,t)=U_0(\xi-\Delta(t))+\sqrt{\epsilon}\Phi(\xi-\Delta(t),t).
\end{equation}
Substituting this decomposition into the neural field equation and carrying out an asymptotic series expansion $\Phi=\Phi_0+O(\epsilon^{1/2})$, one finds that boundedness of $\Phi_0$ leads to a self-consistency condition for $\Delta(t)$, which takes the form of a stochastic ODE. One thus establishes that $\Delta(t)$ undergoes Brownian motion. There have been several subsequent developments of the theory, both in terms of applications and in terms of more rigorous mathematical treatments. Bressloff and Weber have applied these methods to study the effects of noise in binocular rivalry waves, showing that wandering of the wave associated with perceptual switching is diffusive \cite{Webber13}. Second, Kilpatrick and Ermentrout have analyzed the wandering of stationary pulses (bumps) in stochastic neural fields \cite{Kilpatrick13}, and Kilpatrick has shown how weak interlaminar coupling can regularize (reduce the variance) of stationary pulses and propagating waves in a multi-layered, stochastic neural field \cite{Kilpatrick13a,Kilpatrick14}. Regarding rigorous treatments, Faugeras and Inglis \cite{Faugeras14} have addressed the issue of solutions and well-posedness in stochastic neural fields by adapting results from SPDEs, whereas Kruger and Stannat \cite{Kruger14} have developed a rigorous treatment of the multi-scale decomposition of solutions.

One feature that has emerged in some applications of stochastic neural fields is that the displacement variable $\Delta(t)$ can satisfy an Ornstein-Uhlenbeck (OU) process rather than pure Brownian motion. For example, this occurs in the case of stimulus-locked fronts \cite{Bressloff12} and in the regularization of waves in multi-layer networks, where $\Delta(t)$ now represents the relative displacement of fronts in different layers \cite{Kilpatrick14}. However, one assumption in the derivation of the OU process is that the displacement $\Delta(t)$ is small. In this paper, we show that this assumption can break down and, in fact, $\Delta(t)$ evolves according to a nonlinear SDE. It turns out that in the aforementioned applications, the deterministic part of the SDE has a unique stable fixed point, so that one can carry out a linear-noise approximation and recover an OU process, but with modified expressions for the drift term.
More significantly, the nonlinear nature of the SDE also raises the possibility of a breakdown in the linear-noise approximation due to the coexistence of multiple phase-locked states.

Our results are organized as follows. In \S \ref{stimlock}, we demonstrate how a nonlinear SDE can be derived for the stochastic motion of a stimulus-locked front driven by additive noise. The shape of the nonlinearity is determined by both the spatial profile of the stimulus as well as the front profile. We extend our methods in \S \ref{cupfronts} to derive a nonlinear system of SDEs for the motion of two reciprocally coupled fronts in two separate layers, rather than truncating to linear order as in \cite{Kilpatrick13a}. While coupling tends to regularize the propagation of fronts by pulling their positions close together, large deviations in noise can decouple fronts leading to an instability in the case of lateral inhibitory interlaminar coupling (\S \ref{breakdown}). Lastly, we show in \S \ref{cupbumps} that our analysis can be applied to study wandering bumps in reciprocally coupled laminar neural fields. Similar to the case of fronts, the nonlinear system of SDEs we derive can be used to predict the mean waiting time until a large deviation wherein bumps become temporarily decoupled, leading to a phase-slip.

\section{Stimulus-locked fronts with additive noise}
\label{stimlock}

Consider the stochastic neural field equation \cite{Bressloff12}
\begin{eqnarray}
  \tau dU(x,t)&=&\left [-U(x,t)+\int_{-\infty}^{\infty} w(x-y)F(U(y,t))dy\right ]dt\nonumber \\
  && \qquad +\epsilon^{1/2}I(x-vt)dt+\epsilon^{1/2} \tau^{1/2}dW(x,t) ,\quad x\in \Rset
\label{doraI}
\end{eqnarray}
where $I(x-vt)$ represents a moving external stimulus of speed $v$, and $dW(x,t)$ represents an independent Wiener process such that
\begin{equation}
\langle dW(x,t) \rangle =0,\quad \langle  dW(x,t) dW(x',t')\rangle = 2C(x-x')\delta(t-t')dtdt'. 
\label{Wien}
\end{equation}
Here the (stochastic) neural field $U(x,t)$ is a measure of activity within a local population of excitatory neurons at $x\in\Rset$ and time $t$, $\tau$ is a membrane time constant (of order 10 msec), $w(x)$ denotes the spatial distribution of synaptic connections between local populations, and $F(U)$ is a nonlinear firing rate function. $F$ is taken to be a sigmoid function
\begin{equation}
F(u)=\frac{1}{1+\e^{-\gamma(u-\kappa)}}
\end{equation}
with gain $\gamma$ and threshold $\kappa$. In the high--gain limit $\gamma \rightarrow \infty$, this reduces to the Heaviside function
\begin{equation}
F(u)\rightarrow H(u-\kappa)=\left \{ \begin{array}{cc} 1 & \mbox{if $u > \kappa$}, \\
0& \mbox{if $u < \kappa$} \end{array}\right .
\end{equation}
with $H(0)=1/2$.
We will assume that the weight distribution is a positive, even function of $x$, $w(x)\geq 0$ and $w(-x)=w(x)$, and that $w(x)$ is a monotonically decreasing function of $x$ for $x\geq 0$. A common choice is the exponential weight distribution
\begin{equation}
\label{exp}
w(x)=\frac{1}{2\sigma}\e^{-|x|/\sigma},
\end{equation}
where $\sigma$ determines the range of synaptic connections. The latter tends to range from $100\mu$m to $1$ mm \cite{Lund03}.
We fix the units of time by setting $\tau=1$. 

In the absence of any external inputs and noise ($\epsilon =0$), the resulting homogeneous neural field equation 
\begin{eqnarray}
 \frac{\partial u(x,t)}{\partial t}&=&-u(x,t)+\int_{-\infty}^{\infty} w(x-y)F(u(y,t))dy
\label{dora}
\end{eqnarray}
supports a traveling front solution. First note that a homogeneous fixed point solution $u^*$ of equation (\ref{dora}) satisfies
\begin{equation}
u^*=K_0F(u^*),\quad K_0=\int_{-\infty}^{\infty} w(y)dy.
\end{equation}
It can be shown graphically that in the case of a sigmoid rate function, it is possible for the space-clamped network to exhibit bistability, in which a pair of stable fixed points ${u^*}_{\pm}$  are separated by an unstable fixed point $u^*_0$. Let $u(x,t)=U_0(\xi)$, $\xi=x-ct$ be a traveling front solution with wavespeed $c$ such that $\lim_{\xi\rightarrow -\infty}U_0(\xi)=u_+^*$ and $\lim_{\xi\rightarrow \infty}U_0(\xi)=u_-^*$. Substitution of this solution into (\ref{dora}) gives
\begin{eqnarray}
-c U_0'(\xi)&=&-U_0(\xi)+\int_{-\infty}^{\infty} w(\xi-\xi')F(U_0(\xi'))d\xi' .
\label{front}
\end{eqnarray}
Following Amari \cite{Amari77}, an explicit traveling front solution can be constructed in the high gain limit $F(u)\rightarrow H(u-\kappa)$ with $0 < \kappa < K_0$. One finds that $ U_0(\xi)$ is a monotonically decreasing function of $\xi$ that crosses threshold at a unique point. We are free to take the threshold crossing point to be at the origin, since equation (\ref{dora}) is equivariant with respect to uniform translations. That is, $U_0(0) = \kappa$, so that $U_0(\xi) < \kappa $ for $\xi > 0$ and $U_0(\xi) > \kappa$ for
$\xi <0$. Taking $F(u)=H(u-\kappa)$ in equation (\ref{front}) then gives
\begin{eqnarray}
\label{integro-h3}
 -cU_0'(\xi) +U_0(\xi)=\int_{-\infty}^{0}w(\xi-\xi')d\xi'= \int_{\xi}^{\infty}w(x)dx \equiv K(\xi),
\end{eqnarray}
where $U_0'(\xi)=dU_0/d\xi$. Multiplying both sides of the above equation by $\e^{-\xi/c}$ and  integrating with respect to $\xi$ leads to the solution
\begin{equation}
\label{solA}
U_0(\xi)=\e^{\xi/c}\left [\kappa -\frac{1}{c}\int_0^{\xi}\e^{-y/c}{K}(y)dy \right ].
\end{equation}  
Finally, requiring the solution to remain bounded as $\xi \rightarrow \infty$ ($\xi \rightarrow -\infty$) for $c > 0$ (for $c<0$) implies that $\kappa$ must satisfy the condition
\begin{equation}
\label{ck}
\kappa =\frac{1}{|c|}\int_0^{\infty}\e^{-y/|c|}{K}(\mbox{sign}(c)y)dy,
\end{equation}
and thus
\begin{equation}
\label{solA2}
U_0(\xi)=\frac{1}{|c|}\int_{0}^{\infty}\e^{-y/|c|}{K}(\mbox{sign}(c)y+\xi)dy .
\end{equation}  
In the case of the exponential weight distribution (\ref{exp}), the relationship between wavespeed $c$ and threshold $\kappa$ is
\begin{subequations}
\label{ekap}
\begin{align}
c&=c_+(\kappa)\equiv \frac{\sigma}{2\kappa}[1-2 \kappa]\quad \mbox{for} \ \kappa < 0.5, 
\\ c &=c_-(\kappa)\equiv  \frac{\sigma}{2}\frac{1-2\kappa}{1-\kappa} \quad \mbox{for} \ 0.5 < \kappa < 1. 
\end{align}
\end{subequations}
Moreover, one result we will need later is the explicit form for $U_0$ when $\xi>0$ and $c>0$, namely
\begin{equation}
U_0(\xi)= 
\frac{\displaystyle 1}{\displaystyle 2c} \frac{\displaystyle \sigma e^{-\xi/\sigma}   }{\displaystyle 1+ \sigma /c}.  \label{explicitU0}
\end{equation}

In summary, for a Heaviside rate function and no external input there exists a unique stable front solution of equation (\ref{dora}) whose wave speed $c$ is an explicit function of various physiological parameters such as the range of synaptic connections and the firing threshold. Stability of the front can be established by constructing the associated Evans function whose zeros determine the spectrum of the resulting linear operator \cite{Zhang03,Coombes04,Bressloff12a}. If one now includes an external stimulus in the form of a step function of height $I_0$ and speed $v$, one finds that the traveling front can lock to the stimulus provided that $v-c$ lies within a certain range that depends on $I_0$, where $c$ is the natural wave speed in the absence of inputs \cite{Folias05}.
Although it is not possible to obtain explicit traveling wave solutions in the case of a smooth firing rate function, it is possible to prove the existence of a unique front with or without an external stimulus \cite{Ermentrout93,Ermentrout10}.
In this paper, we develop our analysis of wandering fronts in stochastic neural fields using a smooth sigmoid rate function $F$. However, we illustrate the theory by taking the high-gain limit, since this yields explicit formulae for the nonlinear Langevin equation describing wandering in the weak noise limit.

\subsection{Stimulus-locking in the presence of noise} We want to determine from equation (\ref{doraI}) how the combination of a weak moving stimulus and weak additive noise affects the propagation of the above traveling front solution in the small $\epsilon$ limit. Suppose that the input is given by a positive, bounded, monotonically decreasing function of amplitude $I_0=I(-\infty)-I(\infty)$. From the theory of stimulus-locked fronts in deterministic neural fields \cite{Folias05,Webber13}, we expect that in the absence of noise, the resulting inhomogeneous neural field equation can support a traveling front that locks to the stimulus, provided that the stimulus speed $v$ is sufficiently close to the natural speed $c$ of spontaneous fronts, that is,
\[v=c+\sqrt{\epsilon}v_1.\]
On the other hand, following recent studies of wandering fronts \cite{Bressloff12,Webber13} and bumps \cite{Kilpatrick13,Kilpatrick13a,Kilpatrick14} in stochastic neural fields, we expect the additive noise term in equation (\ref{doraI}) to generate two distinct phenomena that occur on different time--scales: a slow stochastic displacement of the front, and fast fluctuations in the front profile.
Both stimulus-locking and stochastic wandering can thus be captured by decomposing the solution $U(x,t)$ of equation (\ref{doraI}) as a combination of a fixed wave profile $U_0$ that is displaced by an amount $\widehat{\Delta}(t)=\Delta(t)+(v-c)t$ from its uniformly translating mean position $\xi=x-ct$, and a time--dependent fluctuation $\Phi$ in the front shape about the instantaneous position of the front:
\begin{equation}
U(x,t)=U_0(\xi-\widehat{\Delta}(t))+\epsilon^{1/2}\Phi(\xi-\widehat{\Delta}(t),t).
\label{UUI}
\end{equation}
Here $U_0$ is the front solution in the absence of inputs moving with natural speed $c$, see equation (\ref{integro-h3}). (For ease of notation, we have suppressed the fact that $\widehat{\Delta}$ and $\Phi$ also depend on $\epsilon$.) It is important to note that as it stands, the decomposition (\ref{UUI}) is non-unique with regards the separate functions $\widehat{\Delta}$ and $\Phi$. Hence, from a functional analytic perspective, one has to impose an additional constraint involving, for example, the difference $\|U(\cdot,t)-U_0(\cdot-\widehat{\Delta}(t)\|$ with respect to an appropriate norm \cite{Kruger14}. In the case of the formal perturbation method developed below, we determine $\widehat{\Delta}(t)$ uniquely to leading order in $\epsilon$ by applying the Fredholm alternative, whereas the leading order expression for $\Phi$ is non-unique. The same comments apply to the decompositions carried out in subsequent sections.

The next step is to substitute the decomposition (\ref{UUI}) into equation (\ref{doraI})\footnote{In carrying out the change of stochastic variables, care has to be taken with regards Ito calcuus \cite{Gardiner09}. That is, it turns out that $\widehat{\Delta}(t)$ evolves according to a diffusion process, which means there will be an Ito-correction term involving the quadratic variation of $\widehat{\Delta}$. However, this term will be of order $\epsilon$ which does not contribute to the leading order Langevin equation for $\widehat{\Delta}$. Therefore, we ignore any Ito-correction terms in our subsequent analysis.}:
\begin{eqnarray}
&&-vU_0'(\xi-\widehat{\Delta}(t))dt-U_0'(\xi-\widehat{\Delta}(t))d\Delta (t)+\epsilon^{1/2}\left [d\Phi(\xi-\widehat{\Delta}(t),t)-v\Phi'(\xi-\widehat{\Delta}(t),t)dt\right ]\nonumber  \\
&& \quad -\epsilon^{1/2}\Phi'(\xi-\widehat{\Delta}(t),t)d\Delta (t)\nonumber  \\
\label{luca}
&& =-U_0(\xi-\widehat{\Delta}(t))dt-\epsilon^{1/2}\Phi(\xi-\widehat{\Delta}(t),t)dt +\epsilon^{1/2}d{W}(\xi+ct,t) \nonumber  \\
&& \quad +\int_{-\infty}^{\infty}w(\xi-\xi')F\left [U_0(\xi'-\widehat{\Delta}(t))+\epsilon^{1/2}\Phi(\xi'-\widehat{\Delta}(t),t)\right ]d\xi'dt \nonumber \\
&&\quad +\epsilon^{1/2} I(\xi-(v-c)t,t). 
\end{eqnarray}
An important point to emphasize is that one cannot generally carry out a perturbation expansion with respect to $\Delta(t)$, as was previously assumed in \cite{Bressloff12}. As we will show below, $\Delta(t) \rightarrow \xi_0$ as $t\rightarrow \infty$ in the absence of noise with the constant $\xi_0$ typically of $O(1)$. On the other hand, one can carry out a perturbation expansion in $\Phi$ by writing $\Phi=\Phi_0+\sqrt{\epsilon}\Phi_1+O(\epsilon)$. Self-consistency of this asymptotic expansion will then determine $\Delta(t)$. Substituting the series expansion for $\Phi$ into equation (\ref{luca}), Taylor expanding the nonlinear function $F$, and imposing the homogeneous equation for $U_0$ leads to the following equation for $\Phi_0$: 
\begin{eqnarray*}
&&-{\epsilon}^{1/2}v_1U_0'(\xi-\widehat{\Delta}(t))dt-U_0'(\xi-\widehat{\Delta}(t))d\Delta (t)\\
&& \quad +\epsilon^{1/2}\left [d\Phi_0(\xi-\widehat{\Delta}(t),t)-c\Phi_0'(\xi-\widehat{\Delta}(t),t)dt\right ]-\epsilon^{1/2}\Phi_0'(\xi-\widehat{\Delta}(t),t)d\Delta (t) \\
&& \quad =-\epsilon^{1/2}\Phi_0(\xi-\widehat{\Delta}(t),t)dt +\epsilon^{1/2}d{W}(\xi+ct,t) +\epsilon^{1/2} I(\xi-(v-c)t,t)
\\
&& \qquad +\epsilon^{1/2}\int_{-\infty}^{\infty}w(\xi-\xi')F'(U_0(\xi'-\widehat{\Delta}(t)))\Phi_0(\xi'-\widehat{\Delta}(t),t)d\xi'dt.
\end{eqnarray*}
Shifting $\xi\rightarrow \xi+\widehat{\Delta}(t)$ and dividing through by $\epsilon^{1/2}$ then gives
\begin{eqnarray}
d\Phi_0(\xi,t) &=&\widehat{L} \Phi_0(\xi,t)dt+\epsilon^{-1/2}U_0'(\xi)d\Delta(t)+d\widetilde{W}(\xi,t)+I(\xi+\Delta(t))dt \nonumber \\
&&+v_1U_0'(\xi)dt,
\label{Phi0}
\end{eqnarray}
where $\widetilde{W}(\xi,t)={W}(\xi+\Delta(t)+vt,t)$ and $\widehat{L}$ is the non--self--adjoint linear operator
\begin{eqnarray}
\widehat{L}A(\xi)= c\frac{d A(\xi)}{d\xi}-A(\xi) +\int_{-\infty}^{\infty} w(\xi-\xi')F'(U_0(\xi'))A(\xi')d\xi'
\label{ve2}
\end{eqnarray}
for any function $A(\xi)\in L^2(\Rset)$. 
It can be shown that for a sigmoid firing rate function and exponential weight distribution, the operator $\widehat{L}$ has a 1D null space spanned by $U_0'(\xi)$ \cite{Ermentrout93}. (The fact that $U_0'(\xi)$ belongs to the null space follows immediately from differentiating equation (\ref{front}) with respect to $\xi$.) We then have the solvability condition for the existence of a bounded solution of equation (\ref{Phi0}), namely, that the inhomogeneous part is orthogonal to all elements of the null space of the adjoint operator $\widehat{L}^*$. The latter is
defined with respect to the inner product
\begin{equation}
\int_{-\infty}^{\infty}B(\xi){\widehat L}A(\xi)d\xi = \int_{-\infty}^{\infty} \left [{\widehat L}^*B(\xi)\right ]A(\xi)d\xi
\end{equation}
where $A(\xi)$ and $B(\xi)$ are arbitrary integrable functions. Hence,
\begin{equation}
\label{L*}
{\widehat L}^*B(\xi)= -c\frac{d B(\xi)}{d \xi} 
-B(\xi)+F'(U_0(\xi))\int_{-\infty}^{\infty}w(\xi-\xi')B(\xi')d\xi'.
\end{equation}
It can be proven that $\widehat{L}^*$ also has a one-dimensional null-space \cite{Ermentrout93}, that is, it is spanned by
some function ${\mathcal V}(\xi)$. The solvability condition reflects the fact that the homogeneous system ($\epsilon=0$) is marginally stable with respect to uniform translations of a front. This means that the linear operator $\widehat{L}$ has a simple zero eigenvalue whilst the remainder of the spectrum lies in the left-half complex plane. Hence, perturbations of $\Phi_0$ that lie in the null-space will not be damped and thus $\Phi_0$ will be unbounded in the large $t$ limit unless these perturbations vanish identically.

Taking the inner product of both sides of equation (\ref{Phi0}) with respect to ${\mathcal V}(\xi)$ leads to the solvability condition
\begin{equation}
\int_{-\infty}^{\infty} {\mathcal V}(\xi)\left [\epsilon^{-1/2}U_0'(\xi)d\Delta(t)+I(\xi+\Delta(t))dt+v_1U_0'(\xi)dt+d\widetilde{W}(\xi,t)\right ]d\xi=0.
\end{equation}
It follows that, to leading order, $\Delta(t)$ satisfies the nonlinear SDE
\begin{equation}
d\Delta(t)+\epsilon^{1/2}G(\Delta(t))dt =  \epsilon^{1/2}d\widehat{W}(t),
\label{SODE}
\end{equation}
where
\begin{equation}
G(\Delta)=
\frac{\displaystyle \int_{-\infty}^{\infty} {\mathcal V}(\xi)[I(\xi+\Delta)+v_1U_0'(\xi)]d\xi}
{\displaystyle  \int_{-\infty}^{\infty} {\mathcal V}(\xi)U_0'(\xi)d\xi}.
\end{equation}
and
\begin{equation}
\widehat{W}(t)=-\frac{\displaystyle \int_{-\infty}^{\infty} {\mathcal V}(\xi)\widetilde{W}(\xi,t)d\xi}
{\displaystyle  \int_{-\infty}^{\infty} {\mathcal V}(\xi)U_0'(\xi)d\xi}.
\end{equation}
Note that 
\begin{equation}
\langle d\widehat{W}(t)\rangle = 0,\quad \langle d\widehat{W}(t)d\widehat{W}(t')\rangle=2 D \delta(t-t')dt'dt
\end{equation}
with $D$ the effective diffusivity
\begin{eqnarray}
D&=&\frac{\displaystyle \int_{-\infty}^{\infty}\int_{-\infty}^{\infty} {\mathcal V}(\xi) {\mathcal V}(\xi')
\langle d\widetilde{W}(\xi,t)d\widetilde{W}(\xi',t)\rangle d\xi d\xi'}
{\displaystyle \left [ \int_{-\infty}^{\infty} {\mathcal V}(\xi)U_0'(\xi)d\xi \right ]^2}\nonumber \\
&=& \frac{\displaystyle \int_{- \infty}^{\infty} \int_{-\infty}^{\infty} {\mathcal V}(\xi)C(\xi-\xi'){\mathcal V}(\xi') d\xi'd\xi}
{\displaystyle  \left [\int_{-\infty}^{\infty} {\mathcal V}(\xi)U_0'(\xi)d\xi \right ]^2}.
\label{D}
\end{eqnarray}
Suppose that there exists a unique $\Delta=\xi_0$ for which $G(\xi_0)=0$ and $G'(\xi_0)>0$. This represents a stable stimulus-locked front in the absence of noise, with $\xi_0$ the relative shift of the stimulus-locked front and the input. Taylor expanding about this solution by setting $\epsilon^{1/2}Y(t)=\Delta(t)-\xi_0$ with $Y(t)=O(1)$ we obtain the OU process\footnote{Note that the $\epsilon$ expansion of the Langevin equation (\ref{SODE}) under the linear noise approximation is distinct from the original $\epsilon$ expansion of the neural field equation. That is, the Langevin equation is an exact self-consistency condition for the phase shift $\Delta(t)$ obtained by applying the Fredholm alternative to the leading order terms in the $\epsilon$ expansion of the neural field equation, see equation (\ref{Phi0}). The corresponding solution $\Phi_0$ of (\ref{Phi0}) is non-unique, since it can be shifted by $A_0U_0'(\xi)$ with $A_0$ arbitrary. In order to determine $A_0$, it is necessary to consider the self-consistency condition obtained by applying the Fredholm alternative to the $O(\epsilon)$ equation for $\Phi_1$. However, this does not modify the equation for $\Delta(t)$.}
\begin{equation}
dY(t)+\epsilon^{1/2}A Y (t)dt=  d\widehat{W}(t),
\label{dY}
\end{equation}
where
\begin{equation*}
A=G'(\xi_0)=
\frac{\displaystyle \int_{-\infty}^{\infty} {\mathcal V}(\xi)I'(\xi+\xi_0)d\xi}
{\displaystyle  \int_{-\infty}^{\infty} {\mathcal V}(\xi)U_0'(\xi)d\xi}
\end{equation*}
(In our previous work \cite{Bressloff12}, $\xi_0$ was assumed to be zero.)
Using standard properties of an Ornstein--Uhlenbeck process \cite{Gardiner09}
\begin{subequations}
\label{OUD}
\begin{align}
\langle \Delta(t)\rangle &=\xi_0\left [1-\e^{-\sqrt{\epsilon}A t}\right ]+ \Delta(0)\e^{-\sqrt{\epsilon}A t},\\ \langle \Delta(t)^2\rangle -\langle \Delta(t)\rangle^2&= \frac{\sqrt{\epsilon} D}{A}\left [1-\e^{-2\sqrt{\epsilon}At}\right ].
\end{align}
\end{subequations}
In particular, the variance approaches a constant $\sqrt{\epsilon} D/A$ in the large $t$ limit and the mean converges to the fixed point $\xi_0$.

In summary, it is necessary to modify the analysis of stimulus-locked fronts in \cite{Bressloff12} by noting that $\Delta(t)$ actually evolves according to the nonlinear SDE (\ref{SODE}) rather than an OU process. In order to obtain an OU process one then has to model Gaussian-like fluctuations about a stable fixed point $\xi_0$ of the deterministic part using  a linear-noise approximation. This is reasonable provided that the deterministic equation does not exhibit multistability, which holds for stimulus-locked fronts. One additional feature of the nonlinear theory is that
an explicit condition for the existence of a stimulus-locked front can be obtained in terms of the fixed point $\xi_0$.
\subsection{Explicit results for Heaviside rate function}

We now illustrate the above analysis by taking $F(u)=H(u-\kappa)$ so that the null vector ${\mathcal V}$ can be calculated explicitly. For the sake of illustration, we also assume that $c>0$. It follows that ${\mathcal V}$ satisfies the equation
\begin{equation}
\label{adj}
 c{\mathcal V}'(\xi)
+{\mathcal V}(\xi)=-\frac{\delta(\xi)}{U_0'(0)}\int_{-\infty}^{\infty}w(\xi'){\mathcal V}(\xi')d\xi',
\end{equation}
which has the solution \cite{Bressloff01}
\begin{equation}
\label{sol-U}
{\mathcal V}(\xi)= -H(\xi)\exp \left (-\xi/c \right ) .
\end{equation}
We have used equation (\ref{solA2}) for $U_0$, which implies that
\begin{equation}
U_0'(\xi)=-\frac{1}{c}\int_0^{\infty}\e^{-y/c}w(y+\xi)dy.
\label{U0'}
\end{equation}
For explicit calculations, we take $w$ to be the exponential weight function, so that $U_0$ has the explicit form (\ref{explicitU0}). In anticipation of subsequent sections, we evaluate equation (\ref{D}) in the case of the correlation function $C( \xi ) = C_0 \cos (\xi)$, for which
\begin{align}  \label{Decos}
D = C_0 \frac{\left[ \int_0^{\infty} \e^{- \xi / c} \cos \xi d \xi \right]^2 + \left[ \int_0^{\infty} \e^{- \xi / c} \sin \xi d \xi \right]^2}{\left[ \int_0^{\infty} \e^{-\xi/c} U_0'( \xi ) d \xi \right]^2} = \frac{C_0 \sigma^2}{4 \kappa^4 (c^2+1)}.
\end{align}
In order to determine the constant shift $\xi_0$ and the drift term $A$, we need to specify the form of the input $I$. For the sake of illustration, let
\[I(\xi)=I_0H(-\xi).\]
The nonlinear function $G(\Delta)$ becomes
\begin{align}
G(\Delta)&=
\frac{\displaystyle \int_{0}^{\infty}I_0 \e^{-\xi/c}[H(-\xi-\Delta)+v_1U_0'(\xi)]d\xi}
{\displaystyle  \int_{0}^{\infty} \e^{-\xi/c}U_0'(\xi)d\xi}=v_1+I_0H(-\Delta)\frac{\displaystyle  \int_0^{-\Delta}e^{-\xi/c}d\xi}{\displaystyle  \int_{0}^{\infty}\e^{-\xi/c}U_0'(\xi)d\xi}
\nonumber \\
&=v_1-I_0 H(-\Delta)\frac{2(c+\sigma)^2}{\sigma}\left [1-\e^{\Delta/c}\right ].
\end{align}

\begin{figure}[t!]                                                    
\centering
\includegraphics[width=12cm,angle=0]{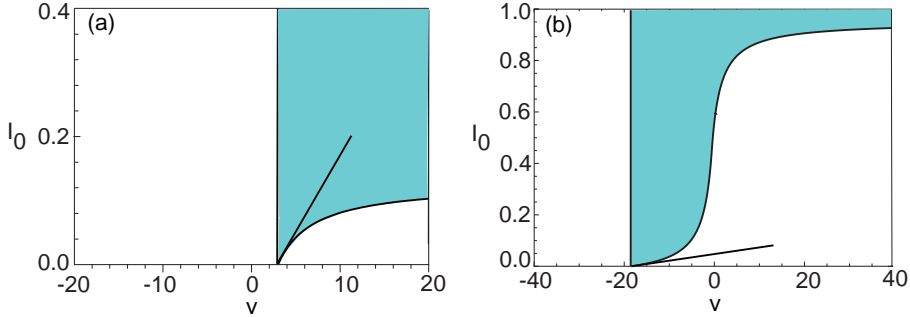}
\caption{\small (a) Existence regions for stimulus-locked traveling fronts in the $(I_0,v)$--plane for $\epsilon =1$, $\sigma=2$ and $c=3$. For a given input amplitude $I_0$, stimulus locking occurs for a finite range of $v$ with $c < v <v^*$. The curved boundary on the right-hand side of the existence region yields a function $I_0=I_0(v^*)$ whose tangent in the limit $I_0\rightarrow 0$ is given by the straight line $I_0=\sigma(v^*-c)\sigma/2(c+\sigma)^2)$. (b) Corresponding existence region for $c=-18$.}
\label{tongue}
\end{figure}

It follows that for a given stimulus velocity there exists a unique stimulus-locked front with shift $\xi_0$ satisfying $G(\xi_0)=0$, that is,
\begin{equation}
\label{xi00}
\xi_0=c\ln\left [1-\frac{\sigma v_1}{2I_0(c+\sigma)^2}\right ] <0,
\end{equation}
provided
\begin{equation}
0<v_1<v_1^*\equiv \frac{2I_0(c+\sigma)^2}{\sigma}.
\end{equation}
Note that equation (\ref{xi00}) is consistent with the non-perturbative result obtained in \cite{Folias05} after Taylor expanding to lowest order in $\epsilon$.
Moreover,
\[G'(\xi_0)=c^{-1}I_0\frac{2(c+\sigma)^2}{\sigma}\e^{\xi_0/c}=\frac{v_1}{c}\frac{\e^{\xi_0/c}}{1-\e^{\xi_0/c}} > 0,\]
so the fixed point is stable. In Fig. \ref{tongue}(a) we plot an example of an existence region for stimulus-locked fronts in the $(v,I_0)$-plane and $c>0$. In the limit of small inputs the tangent to the existence curve approaches the straight line obtained using perturbation theory. An analogous calculation can be carried out for $c<0$ - an example existence region is shown in Fig. \ref{tongue}(b). The stationary variance about the fixed point takes the explicit form
\begin{equation}
\langle \Delta(t)^2\rangle -\langle \Delta(t)\rangle^2=\frac{\sqrt{\epsilon}D}{G'(\xi_0)}.
\end{equation}
Hence, increasing the stimulus velocity reduces the variance. Note that the singularity in the limit $v\rightarrow c$ reflects the fact that the stimulus-locked front becomes unstable.

\section{Fronts in weakly-coupled stochastic neural fields}
\label{cupfronts}

Another example of wandering fronts exhibiting an OU process occurs in a laminar neural field model \cite{Kilpatrick14}. As in the previous example of stimulus-locked fronts, we will show that the noise-induced wandering is described by a nonlinear SDE, and it is only by linearizing about a stable fixed point of the SDE that one obtains an OU process. For the sake of illustration, consider a pair of identical 1D neural fields labeled $j=1,2$ that are mutually coupled via the interlaminar weight distributions $J_1(x)$ and $J_2(x)$:
\begin{subequations}
\begin{eqnarray}
\label{FT1}
  \tau dU_1(x,t)&=&\left [-U_1(x,t)+\int_{-\infty}^{\infty} w(x-y)F(U_1(y,t))dy\right ]dt\\
  && \qquad +\epsilon^{1/2}\int_{-\infty}^{\infty} J_1(x-y)F(U_2(y,t))dydt+\epsilon^{1/2} \tau^{1/2}dW_1(x,t) ,\nonumber 
\end{eqnarray}
and
\begin{eqnarray}
\label{FT2}
  \tau dU_2(x,t)&=&\left [-U_2(x,t)+\int_{-\infty}^{\infty} w(x-y)F(U_2(y,t))dy\right ]dt\\
  && \qquad +\epsilon^{1/2}\int_{-\infty}^{\infty} J_2(x-y)F(U_1(y,t))dydt+\epsilon^{1/2} \tau^{1/2}dW_2(x,t) .\nonumber 
\end{eqnarray}
\end{subequations}
Here $W_1(x,t)$ and $W_2(x,t)$ represent spatially extended Wiener processes such that
\begin{equation}
\langle dW_j(x,t) \rangle =0,\quad \langle  dW_i(x,t) dW_j(x',t')\rangle = 2C_{ij}(x-x')\delta(t-t')dtdt'. 
\label{Wien2}
\end{equation}
Note that the interlaminar coupling is assumed to be weak and asymmetric (unless $J_1=J_2$). As with the intralaminar coupling, the distributions $J_j(x)$ will be taken to be positive exponentials:
\begin{equation}
J_j(x)={\alpha_j}\e^{-|x|/\sigma_j}.  \label{cupexp}
\end{equation}
As shown by Kilpatrick \cite{Kilpatrick14}, in the absence of noise, the interlaminar coupling phase-locks the fronts propagating in each of the two networks, resulting in a composite front with fixed relative shift $\xi_0$. We wish to derive conditions for such locking and determine how the presence of noise induces wandering of the composite front relative to $\xi_0$.

\subsection{Interlaminar coupling of fronts in the presence of noise}
In the absence of interlaminar coupling and noise $(\epsilon=0$), each neural field independently supports a traveling front solution $U_0(\xi)$ along the lines outlined in \S 2 with $\xi=x-ct$. Since each neural field is homogeneous the threshold crossing point of each front is arbitrary. Following our analysis of stimulus-locked fronts, we can simultaneously investigate the effects of weak coupling and noise by considering the decompositions
\begin{equation}
U_j(x,t)=U_0(\xi-{\Delta}_j(t))+\epsilon^{1/2}\Phi_j(\xi-{\Delta}_j(t),t),\quad j=1,2
\label{U}
\end{equation}
with  $\Phi_j=\Phi_{j0}+\sqrt{\epsilon}\Phi_{j1}+O(\epsilon)$.
Substituting into equations (\ref{FT1}) and (\ref{FT2}), Taylor expanding the nonlinear function $F$, and imposing the homogeneous equation (\ref{solA}) for $U_0$ leads to the following equations for $\Phi_{j0}$: 
\begin{subequations}
\begin{eqnarray}
d\Phi_{10}(\xi,t) &=&\widehat{L} \Phi_{10}(\xi,t)dt+\epsilon^{-1/2}U_0'(\xi)d\Delta_1(t)+d\widetilde{W}_1(\xi,t)\nonumber \\
&&+\int_{-\infty}^{\infty}J_1(\xi-\xi')F(U_0(\xi'+\Delta_1(t)-{\Delta_2}(t)))d\xi'dt
\label{p1}
\end{eqnarray}
and
\begin{eqnarray}
d\Phi_{20}(\xi,t) &=&\widehat{L} \Phi_{20}(\xi,t)dt+\epsilon^{-1/2}U_0'(\xi)d\Delta_2(t)+d\widetilde{W}_2(\xi,t)\nonumber \\
&&+\int_{-\infty}^{\infty}J_2(\xi-\xi')F(U_0(\xi'+\Delta_2(t)-{\Delta_1}(t)))d\xi'dt,
\label{p2}
\end{eqnarray}
\end{subequations}
where
\[\widetilde{W}_j(\xi,t)=W_j(\xi+ct+{\Delta_j}(t),t),\]
and the linear operator $\widehat{L}$ is given by equation (\ref{ve2}).

As in the case of stimulus-locked fronts, boundedness of the fast fluctuations $\Phi_{j0}$ require that the inhomogeneous terms are orthogonal to the null vector ${\mathcal V}(\xi)$ of the adjoint operator $\widehat{L}^*$. This leads to the following two-dimensional nonlinear SDE for $(\Delta_1(t),\Delta_2(t))$:
\begin{subequations}
\label{SODE2}
\begin{align}
d\Delta_1(t)+\epsilon^{1/2}G_1(\Delta_1(t)-\Delta_2(t))dt &=  \epsilon^{1/2}d\widehat{W}_1(t),\\
d\Delta_2(t)+\epsilon^{1/2}G_2(\Delta_2(t)-\Delta_1(t))dt &=  \epsilon^{1/2}d\widehat{W}_2(t)
\end{align}
\end{subequations}
where
\begin{equation}
\label{lamG}
G_j(\Delta)=
\frac{\displaystyle \int_{-\infty}^{\infty} {\mathcal V}(\xi) \left [\int_{-\infty}^{\infty}J_j(\xi-\xi')F(U_0(\xi'+\Delta)d\xi'\right ]d\xi}
{\displaystyle  \int_{-\infty}^{\infty} {\mathcal V}(\xi)U_0'(\xi)d\xi}.
\end{equation}
and
\begin{equation}
\label{gogo}
\widehat{W}_j(t)=-\frac{\displaystyle \int_{-\infty}^{\infty} {\mathcal V}(\xi)\widetilde{W}_j(\xi,t)d\xi}
{\displaystyle  \int_{-\infty}^{\infty} {\mathcal V}(\xi)U_0'(\xi)d\xi}.
\end{equation}
Note that 
\begin{equation}
\langle d\widehat{W}_i(t)\rangle = 0,\quad \langle d\widehat{W}_j(t')d\widehat{W}_k(t)\rangle=2 D_{jk} \delta(t-t')dt'dt
\end{equation}
with $D_{jk}$ the effective diffusion matrix
\begin{eqnarray}
D_{jk}&=&\frac{\displaystyle \int_{-\infty}^{\infty}\int_{-\infty}^{\infty} {\mathcal V}(\xi) {\mathcal V}(\xi')
\langle d{W}_j(\xi+ct+\Delta_j(t),t)d{W}_k(\xi'+ct+\Delta_k(t),t)\rangle d\xi d\xi'}
{\displaystyle \left [ \int_{-\infty}^{\infty} {\mathcal V}(\xi)U_0'(\xi)d\xi \right ]^2}\nonumber \\
&=& \frac{\displaystyle \int_{-\infty}^{\infty} {\mathcal V}(\xi)C_{jk}(\xi-\xi'+\Delta_j(t)-\Delta_k(t)){\mathcal V}(\xi') d\xi'd\xi}
{\displaystyle  \left [\int_{-\infty}^{\infty} {\mathcal V}(\xi)U_0'(\xi)d\xi \right ]^2}.
\label{D2}
\end{eqnarray}
It is clear that if we allow correlations between the noise in different layers, then the diffusion matrix depends on the phase difference $\Delta_1-\Delta_2$, and the noise terms in the SDE (\ref{SODE2}) are multiplicative. In order to avoid issues regarding the interpretation of the noise in terms of Ito or Stratonovich, we will simplify the correlation matrix by setting
\[C_{ij}(\xi)=\delta_{ij}C(\xi).\]
It follows that $D_{jk}=\delta_{jk}D$ with $D$ given by equation (\ref{D}).

In the original analysis of interlaminar neural fields \cite{Kilpatrick14}, $\Delta\equiv \Delta_1-\Delta_2$ was assumed to be small, and equations (\ref{p1}) and (\ref{p2}) were Taylor expanded to first-order in $\Delta$, resulting in a multivariate OU process. However, as we found for stimulus-locked fronts, $\Delta(t)$ is not necessarily small. Therefore, one should proceed by looking for a stable fixed point of the ODE
\begin{equation}
\label{detD}
\frac{d\Delta}{dt}=-\epsilon^{1/2}G_-(\Delta),
\end{equation}
which is obtained by subtracting the deterministic parts of the pair of equations (\ref{SODE2}) and setting $G_{\pm}(\Delta)\equiv G_1(\Delta)\pm G_2(-\Delta)$. 
Suppose that there exists a $\xi_0$ for which $G_-(\xi_0)=0$ and $G_-'(\xi_0)>0$. This represents a stable phase-locked state in the absence of noise, with $\xi_0$ the relative shift of the fronts in the two networks. One can establish the existence of a unique phase-locked state using the properties of $G_j(\Delta)$ given by equation (\ref{lamG}). Since the interlaminar coupling is taken to be excitatory, $J_{j}(\xi)>0$ for all $\xi$, and $U_0(\xi)$ is a monotonically decreasing function of $\xi$, the following properties hold:

\noindent (i) $G_j(\Delta) <0$ for all $\Delta$

\noindent (ii) $G_j(\Delta) $ is a monotonically increasing function of $\Delta$ with $G_j(\Delta)\rightarrow 0$ as $\Delta \rightarrow \infty$
and $G_j(\Delta)\rightarrow \overline{G}_j$ as $\Delta \rightarrow -\infty$

\noindent (iii) The functions $G_1(\Delta) $ and $G_2(-\Delta)$ intersect at a unique point $\Delta=\xi_0$ with the sign of $\xi_0$ determined by the relative strengths of $J_1$ and $J_2$. By symmetry, if $J_1=J_2$ then $\xi_0=0$.

Given the stable fixed point $\xi_0$, one can now apply a linear-noise approximation to the full SDE (\ref{SODE2}) and derive a multivariate OU process. Let
\[\Delta(t)=\xi_0+\epsilon^{1/2}Y(t),\quad S(t)={\Delta_1(t)+\Delta_2(t)}\]
and write
\[\Delta_1(t)=\frac{\xi_0}{2}+\frac{\epsilon^{1/2}}{2}[ S(t)+Y(t)],\quad \Delta_2(t)=-\frac{\xi_0}{2}+\frac{\epsilon^{1/2}}{2}[S(t)-Y(t)].\]
Here $S(t)/2$ represents the ``center-of-mass'' coordinate.
Equations (\ref{SODE2}) become
\begin{subequations}
\label{SODE3}
\begin{align}
\frac{1}{2}[dS(t)+dY(t)]+G_1(\xi_0+\epsilon^{1/2}Y(t))dt &= d\widehat{W}_1(t),\\
\frac{1}{2}[dS(t)-dY(t)]+G_2(-\xi_0-\epsilon^{1/2}Y(t))dt &=  d\widehat{W}_2(t).
\end{align}
\end{subequations}
We can now Taylor expand the nonlinear function $G$ with respect to $\epsilon^{1/2}Y(t)$. Adding and subtracting the above equations then yields the linear system of SDEs
\begin{subequations}
\label{SODE4}
\begin{align}
dS(t)+[G_+(\xi_0)+\epsilon^{1/2}G_+'(\xi_0)Y(t)]dt &=  d{W}_+(t),\\
dY(t)+\epsilon^{1/2}G'_-(\xi_0)Y(t)dt &=  d{W}_-(t),
\end{align}
\end{subequations}
where 
\[{W}_{\pm}(t)=\widehat{W}_1(t)\pm \widehat{W}_2(t).\]
$W_{\pm}(t)$ are also independent Wiener processes with
\[\langle dW_{a}(t)\rangle =0,\quad \langle dW_a(t)dW_b(t)\rangle =4D\delta_{a,b}dt,\quad a,b=\pm .\]
Under the linear-noise approximation, fluctuations in the phase difference $\Delta(t)$ about the fixed point $\xi_0$ satisfy a one-dimensional OU process, which in turn drives fluctuations in the center-of-mass variable $S(t)$. If $\xi_0=0$, then we recover the results of Kilpatrick \cite{Kilpatrick14}. 

It immediately follows that the mean and variance of $Y(t)$ are given by the equations (cf. equation (\ref{OUD}))
\begin{equation}
\bar{Y}(t)\equiv \langle Y(t)\rangle=Y(0) \e^{-\sqrt{\epsilon}A t},\quad \Theta_Y(t)\equiv \langle Y(t)^2\rangle-\langle Y(t)\rangle^2=\frac{2 D}{\sqrt{\epsilon}A}\left [1-\e^{-2\sqrt{\epsilon}At}\right ],
\end{equation}
with $A=G_-'(\xi_0) \equiv G_1'(\xi_0) - G_2'(\xi_0)$ and $G_j$ defined by equation (\ref{lamG}). This implies that in the limit $t\rightarrow \infty$ the mean of the phase-shift $\Delta(t)$ converges to the deterministic fixed point $\xi_0$ and the variance about the fixed point converges to ${2 D\sqrt{\epsilon}}/{A}$. In the case of $S(t)$, 
equation (\ref{SODE4}a) can be integrated explicitly to give (with $S(0)=0$)
\[ S(t)=-G_+(\xi_0)t-\epsilon^{1/2}G_+'(\xi_0)\int_0^tY(s)ds+W_+(t).\]
Taking the mean and variance of this solution yields
\begin{equation}
\bar{S}(t)\equiv \langle S(t)\rangle=-G_+(\xi_0)t-\frac{G_+'(\xi_0)Y(0)}A\left [1-\e^{-\sqrt{\epsilon}At}\right ],
\end{equation}
and
\begin{align}
\Theta_S^2\equiv \langle S^2\rangle-\bar{S}^2&=\epsilon G_+'(\xi_0)^2\int_{0}^t\int_0^t\left [\langle Y(s)Y(s')\rangle -\langle Y(s)\rangle \langle Y(s')\rangle\right ]ds'ds + 4D.
\end{align}
In the large time limit, $Y(t)$ becomes a stationary process with zero mean and autocorrelation function
\[\langle Y(s)Y(s')\rangle=\frac{D}{\sqrt{\epsilon}A}\e^{-\sqrt{\epsilon}A|s-s'|}.\]
Moreover,
\[\int_{0}^t\int_0^t \e^{-\sqrt{\epsilon}A|s-s'|}ds'ds=2\int_{0}^t \int_0^s \e^{-\sqrt{\epsilon}A(s-s')}ds'ds=\frac{2t}{\sqrt{\epsilon}A}-\frac{2}{{\epsilon}A^2}\left [1-\e^{-\sqrt{\epsilon}At}\right ].
\]
Hence, for large $t$
\begin{align}
\Theta_S^2\approx 2Dt,
\end{align}
where we have used the fact that $A=G_-'(\xi_0)$. 

In conclusion, the mean position of each front is given by (for $Y(0)=0$)
\begin{equation}
\langle \Delta_1(t)\rangle =\frac{\xi_0}{2}-\frac{\epsilon^{1/2}G_+(\xi_0)}{2}t\quad \langle \Delta_2(t)\rangle =-\frac{\xi_0}{2}-\frac{\epsilon^{1/2}G_+(\xi_0)}{2}t.
\end{equation}
Thus, as previously noted by Kilpatrick \cite{Kilpatrick14}, one effect of weak coupling is to induce an $O(\sqrt{\epsilon})$ increase in the mean speed of each front according to
\[c\rightarrow c+\frac{\epsilon^{1/2}|G_+(\xi_0)|}{2} > c.\]
Furthermore, in the presence of additive noise, fluctuations in the phase-shift are given by an OU process, whereas fluctuations in the center-of-mass are given by Brownian diffusion in the large-time limit.

\subsection{Explicit results for Heaviside rate function}
Since the uncoupled, deterministic neural field equations ($\epsilon=0$) are given by equation (\ref{dora}), the calculation of the null vector ${\mathcal V}(\xi)$ and diffusivity $D$ for the Heaviside rate function proceeds as in \S 2.2. That is, ${\mathcal V}(\xi)=-H(\xi)\e^{-\xi/c}$. Therefore, we only have to calculate the nonlinear function $G_j(\Delta)$ of equation (\ref{lamG}). First note that
\[\int_{-\infty}^{\infty} {\mathcal V}(\xi)U_0'(\xi)d\xi=-\int_0^{\infty}\e^{-\xi/c}U_0'(\xi)d\xi=\frac{1}{2}\frac{\sigma c}{(\sigma +c)^2}, \]
so that
\[G_j(\Delta)=-2\frac{(\sigma +c)^2}{\sigma c}R_j(\Delta)\]
with
\begin{equation}
R_j(\Delta)=\int_{0}^{\infty} \e^{-\xi/c} \left [\int_{-\infty}^{-\Delta}J_j(\xi-\xi')d\xi'\right ]d\xi.
\end{equation}
Taking $J_j(\xi)$ to be an exponential distribution, we find that for $\Delta >0$
\begin{align*}
R_j(\Delta)&=\frac{c\alpha_j\sigma_j^2}{c+\sigma_j} \e^{-\Delta/\sigma_j}.
\end{align*}
whereas for $\Delta > 0$, 
\begin{align*}
& R_j( - \Delta ) =  \alpha_j c \sigma_j \left( 2 + \frac{\sigma_j \e^{- \Delta/ \sigma_j}}{c- \sigma_j} - \frac{2 c^2 \e^{- \Delta/ c}}{c^2 - \sigma_j^2} \right).
\end{align*}

\begin{figure}[t!]
\begin{center}
\includegraphics[width=8cm]{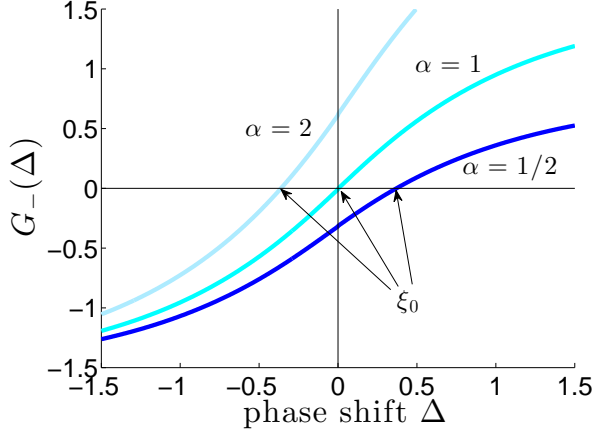}
\caption{\small Plot of $G_-(\Delta)$ as a function of phase shift $\Delta$ for different values of the relative weight $\alpha=\alpha_2/\alpha_1$ of excitatory coupling. Other parameter values are $c=0.25$ and $\sigma=\sigma_1=\sigma_2=1$. It can be seen that there exists a unique zero $\xi_0$ for which $G_-(\xi_0)=0$ and $G'(\xi_0)>0$. If the interlaminar weights are stronger (weaker) in the direction $2\rightarrow 1$ then $\xi_0>0$ ($\xi_0<0$). In the case of symmetric inter laminar connections we have $\xi_0=0$.}
\label{lock}
\end{center}
\end{figure}

\begin{figure}[t!]
\begin{center} \includegraphics[width=6.2cm]{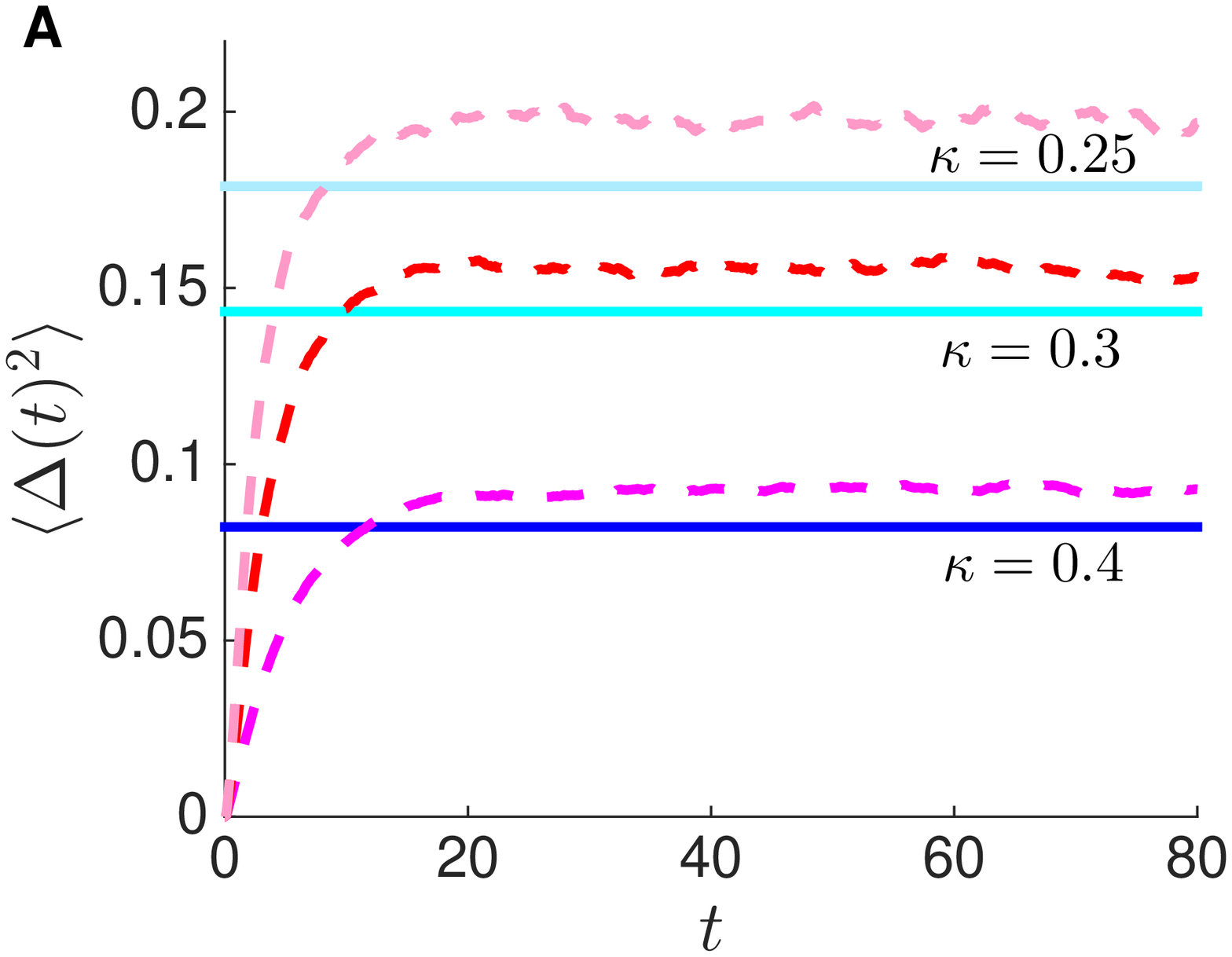} \includegraphics[width=6.2cm]{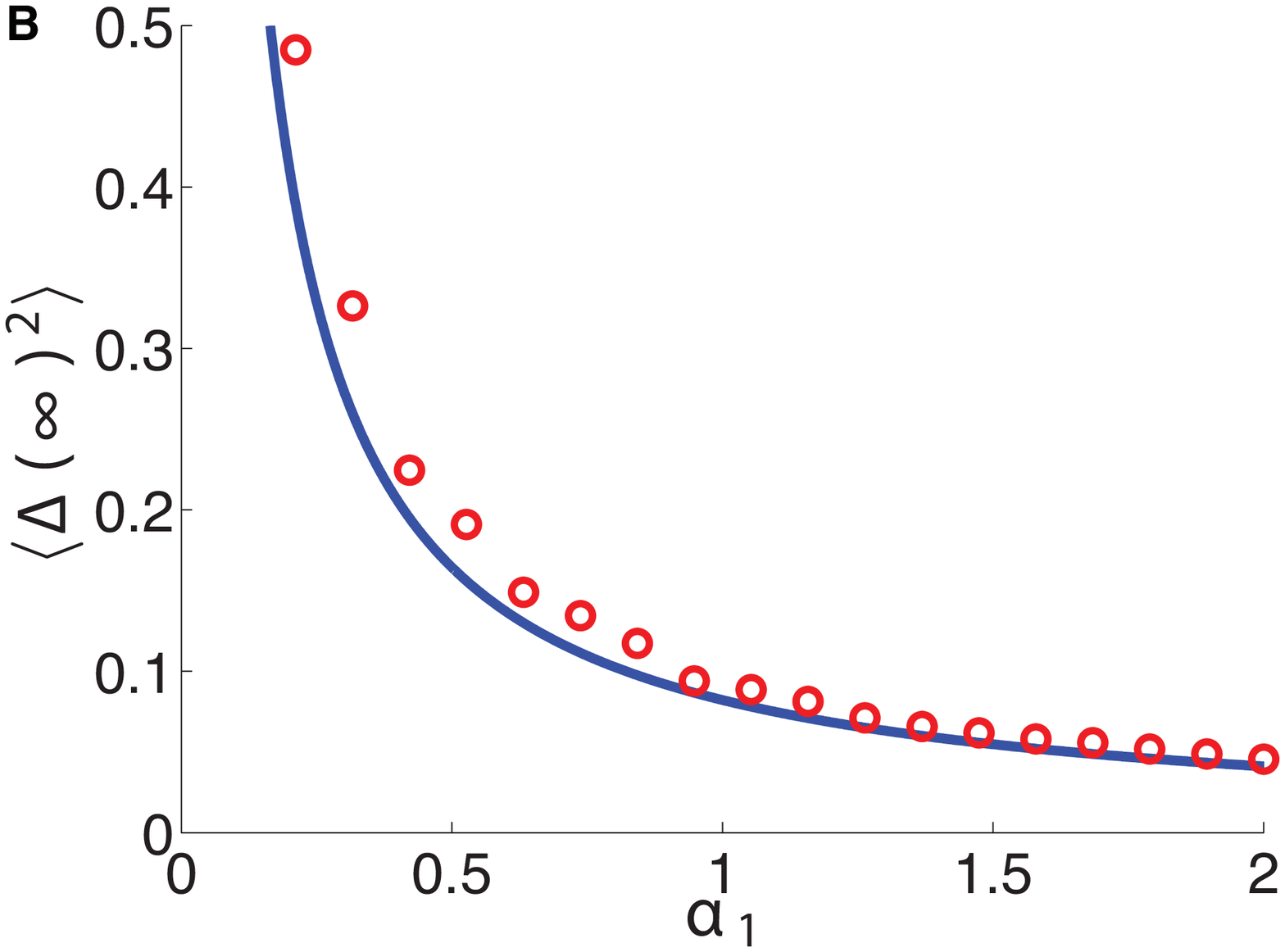} \end{center}
\caption{\small ({\bf A}) The variance in the phase difference $\langle \Delta^2 \rangle$ saturates in the limit $t \to \infty$. Variance is computed as a function of time using numerical simulations (dashed line) and asymptotic theory (solid line) predicts the saturation value (\ref{fvarcos}). Interlaminar connectivity strength $\alpha_1 = \alpha_2 = 1$ so $\xi_0 = 0$ and $\langle \Delta (t) \rangle \equiv 0$; noise amplitude $\epsilon = 0.0005$. ({\bf B}) The stationary variance decreases as a function of interlaminar connectivity strength $\alpha_1 = \alpha_2$ in numerical simulations (circles) and theory (solid line). Threshold $\kappa = 0.4$ and noise amplitude $\epsilon = 0.0005$. Variances are computed using 25000 realizations each.}
\label{fvarplot}
\end{figure}

For the sake of illustration, suppose that $c=0.25$, $\sigma=\sigma_j=1$ and set $\alpha_2/\alpha_1=\alpha$. In Fig. \ref{lock} we plot the function $G_-(\Delta)$ with respect to the phase-shift $\Delta$ for several values of $\alpha_2/\alpha_1$. As expected, $G_-(\Delta)$ is a monotonically increasing function of $\Delta$ with a unique zero $\xi_0$. In the case of symmetric weights ($\alpha=0$) we have $\xi_0=0$, whereas for $0<\alpha <1$ ($1<\alpha$) we find that $\xi_0>0$ ($\xi_0 <0$). Recall that the stationary variance about the phase-locked state is
\begin{equation}
\langle \Delta(t)^2\rangle -\langle \Delta(t)\rangle^2=\frac{2\sqrt{\epsilon}D}{G_-'(\xi_0)}.
\end{equation}
$\xi_0$ is fixed by fixed the degree of asymmetry $\alpha$ in the inter laminar connections. However, $G_-'(\xi_0)$ scales with $\alpha_1$, which establishes that increasing the strength of interlaminar coupling can reduce the variance, as previously highlighted by Kilpatrick \cite{Kilpatrick14}. Upon considering weight spatial scale $\sigma = \sigma_j = 1$, symmetric connectivity $\alpha = 1$, and cosine noise correlations $C(\xi) = \cos (\xi)$ then $D$ is given by (\ref{Decos}), $\xi_0 = 0$, $G_-'(0) = 2 \alpha_1/\kappa$ so $\langle \Delta (t) \rangle \equiv 0$ and 
\begin{align}
\langle \Delta(t)^2\rangle -\langle \Delta(t)\rangle^2=\frac{\sqrt{\epsilon}}{4 \alpha_1 \kappa^3 (c^2+1)}.  \label{fvarcos}
\end{align}
We compare the formula (\ref{fvarcos}) to numerical simulations in Fig. \ref{fvarplot}, demonstrating how the variance decreases with $\alpha_1$ as well as the threshold $\kappa$. 

\section{Breakdown of linear-noise approximation}
\label{breakdown}
\begin{figure}[t!]
\begin{center}
\includegraphics[width=8cm]{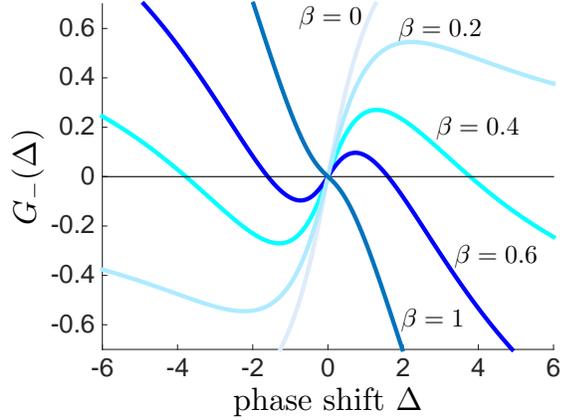}
\caption{\small Plot of $G_-(\Delta)$ as a function of phase-shift $\Delta$ for different values of the inhibitory weight $\beta$. Other parameter values are $\sigma=1$, $c=0.25$, $\alpha_1 = 0.25$, and $\gamma=4$. It can be seen that for intermediate values of $\beta$, a stable phase-locked state at $\Delta =0$ coexists with a pair of unstable fixed points at $\Delta =\pm \xi_0(\beta)$.}
\label{lock2}
\end{center}
\end{figure}

 So far we have considered examples of stochastic phase-locked fronts, in which there is a unique, stable phase-locked state in the deterministic limit. This allowed us to carry out a linear-noise approximation and thus show that fluctuations about the phase-locked state can be characterized in terms of an OU process. Here we consider an example where the linear-noise approximation breaks down due to the existence of an additional unstable phase-locked state in the absence of noise. We again consider a pair of coupled neural fields given by equations (\ref{FT1}) and (\ref{FT2}), but now take the interlaminar coupling to be given by a difference of exponentials: $J_1=J_2=J$ with
 \begin{equation}
 \label{dog}
 J(x)= \alpha_1 \left[ \e^{-|x|}-{\beta}\e^{-|x|/\gamma} \right],\quad 0\leq \beta < 1,\quad \gamma >1.
 \end{equation}
 This distribution represents short-range excitation and long-range inhibition. For the sake of illustration, we assume that the interlaminar coupling is symmetric. The analysis of stochastic phase-locking proceeds along identical lines to \S 3, and we obtain the nonlinear SDE (\ref{SODE2}). In the absence of noise, the phase-difference $\Delta(t)$ evolves according to equation (\ref{detD}) with $G_-(\Delta)=G(\Delta)-G(-\Delta)$ and
\begin{equation}
\label{lamG2}
G(\Delta)\equiv 
\frac{\displaystyle \int_{-\infty}^{\infty} {\mathcal V}(\xi) \left [\int_{-\infty}^{\infty}J(\xi-\xi')F(U_0(\xi'+\Delta)d\xi'\right ]d\xi}
{\displaystyle  \int_{-\infty}^{\infty} {\mathcal V}(\xi)U_0'(\xi)d\xi}
\end{equation} 
for $J$ given by equation (\ref{dog}). In the case of a Heaviside rate function, $G_(\Delta)$ may be evaluated explicitly to yield
\begin{align}
G_-(\Delta)&= 4 \alpha_1 \frac{(\sigma +c)^2}{\sigma} \frac{1}{c^2-1} \left [  c^2 (1 - \e^{- \Delta/c}) - (1 -  \e^{-\Delta}) \right] \\
&\quad - 4 \alpha_1 \beta \gamma \frac{(\sigma +c)^2}{\sigma} \frac{1}{c^2-\gamma^2} \left [  c^2 (1 - \e^{- \Delta/c}) - \gamma^2 (1 -  \e^{-\Delta / \gamma}) \right].  \nonumber
\end{align}
In Fig. \ref{lock2}, we plot $G_-(\Delta)$ for various values of $\beta$ with $\sigma=c=1$ and $\gamma=4$. It can be seen that for a range of $\beta$ values, there exist three fixed points, a stable fixed point at $\Delta=0$ and a pair of unstable fixed points at $\Delta=\pm \xi_1(\beta)$. Note, the lower bound value of $\beta$ at which the unstable fixed points emerge is where
\begin{align}
\lim_{\Delta \to \infty} G_- ( \Delta) = 4 \alpha_1 \frac{(\sigma + c)^2}{\sigma} \left[ \frac{c^2-1}{c^2-1} - \beta \gamma \frac{c^2 - \gamma^2}{c^2 - \gamma^2} \right] \equiv 0,
\end{align}
or where $\beta = \gamma^{-1}$. For sufficiently large $\beta$ (strong mutual inhibition), the system undergoes a pitchfork bifurcation, resulting in a single unstable phase-locked state. This occurs when $G_-'(0) = 0$, equivalently at
\begin{align}
\beta = \beta_{pitch} := \frac{c+ \gamma}{\gamma (c+1)}.
\end{align}

\begin{figure}[tb]
\begin{center} \includegraphics[width=12cm]{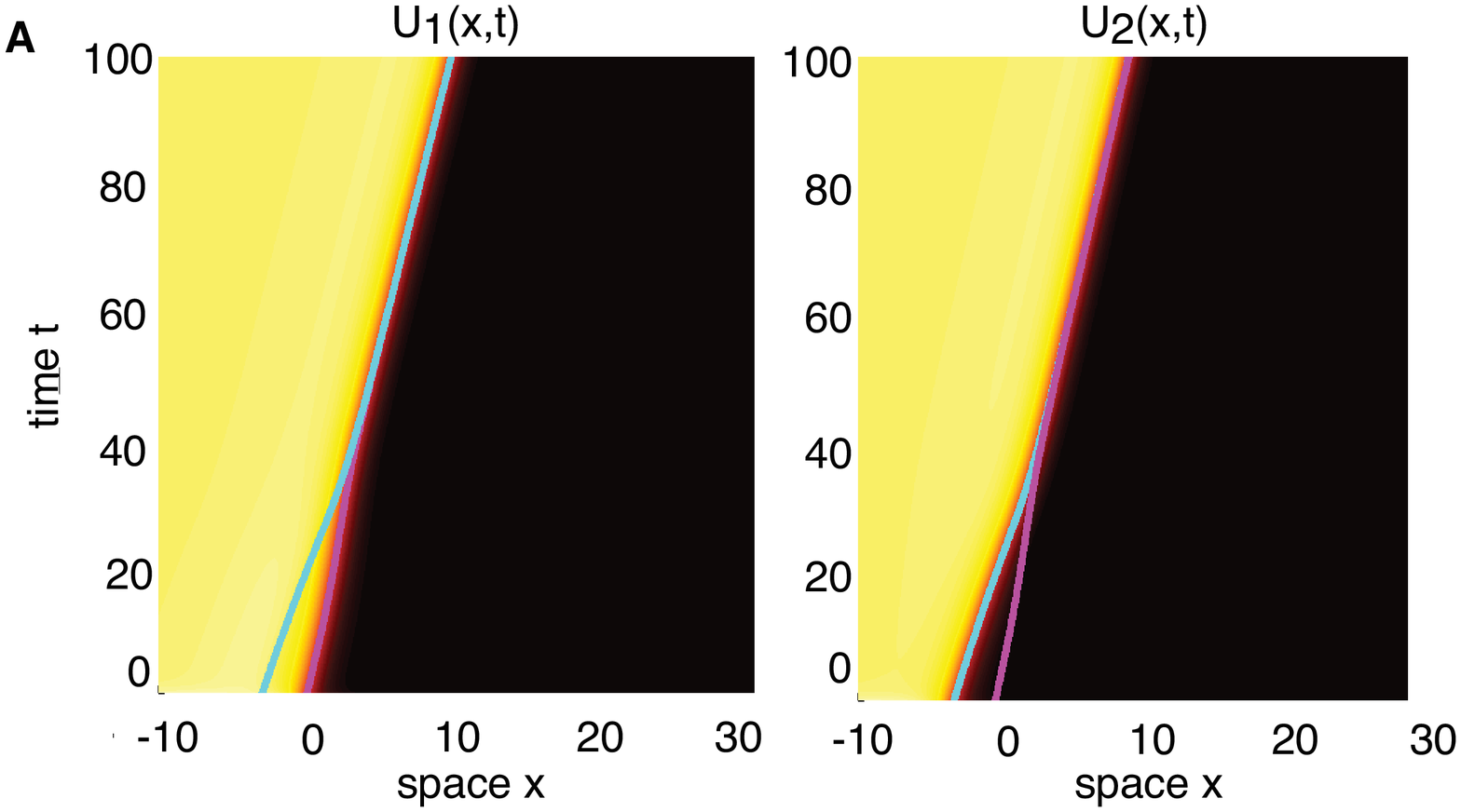} \includegraphics[width=11.5cm]{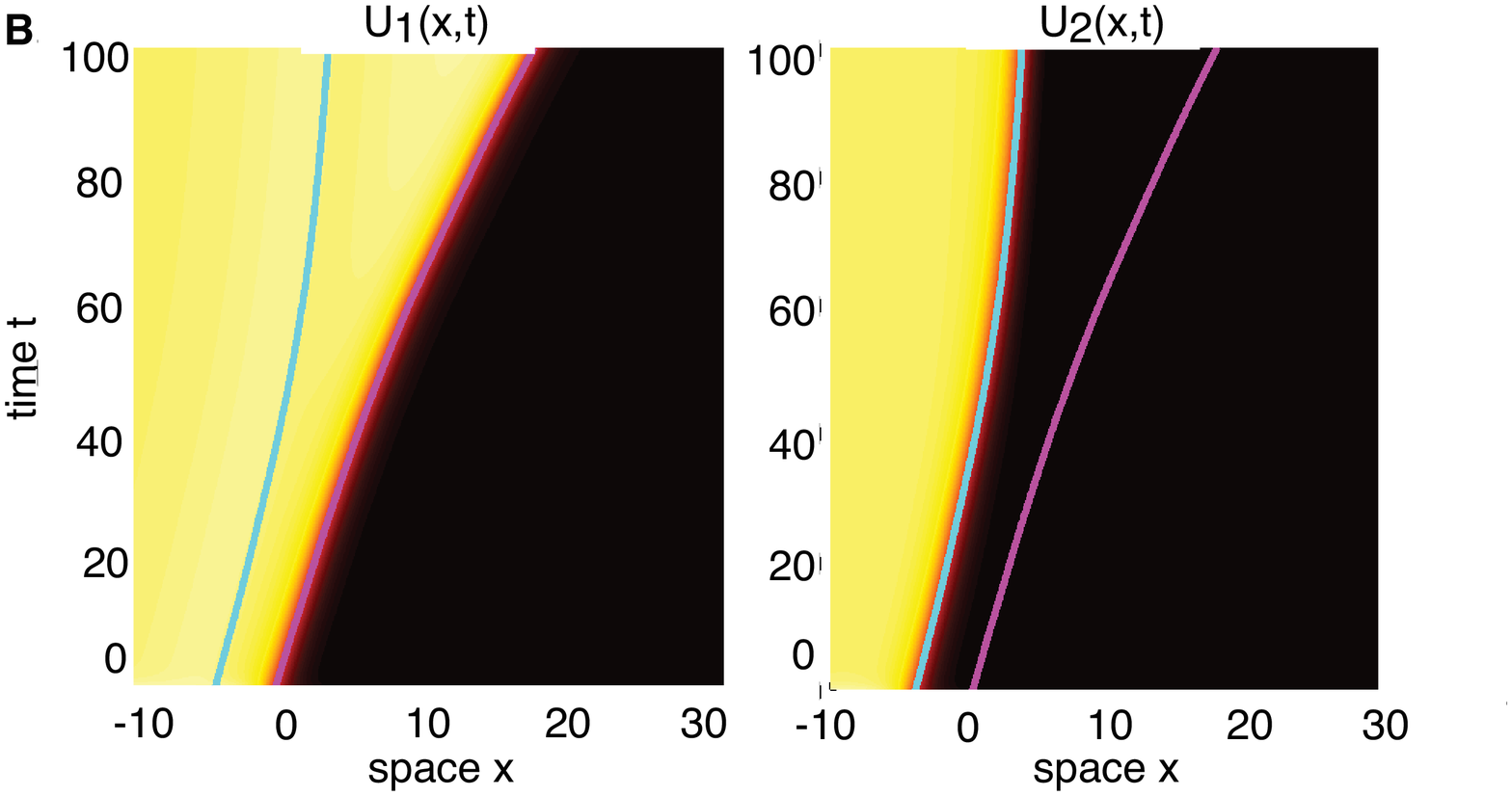} \end{center}
\caption{\small Decoupling of fronts in the case of laterally inhibitory connectivity in (\ref{FT1} - \ref{FT2}) with no noise. ({\bf A}) For a symmetric network, starting the front positions $\Delta_1$ and $\Delta_2$ within the basin of attraction of the coupled state ($\Delta_1 (0) - \Delta_2(0) = 3$) leads to long term coupling: $\lim_{t \to \infty} |\Delta_1(t) - \Delta_2(t)| = \lim_{t \to \infty} | \Delta (t) | = 0$. ({\bf B}) Starting front positions $\Delta_1$ and $\Delta_2$ outside the basin of attraction of the coupled state ($\Delta_1 (0) - \Delta_2(0) = 4$) leads to long term decoupling: $\lim_{t \to \infty} |\Delta_1(t) - \Delta_2(t)| = \lim_{t \to \infty} | \Delta (t) | =  \infty$.  Threshold $\kappa = 0.4$ so $c = 0.25$; no noise ($W_1 = W_2 \equiv 0$); coupling parameters are $\epsilon = 0.005$, $\alpha = 1$, $\beta = 0.4$, $\gamma = 4$. This results in the unstable fixed point (separatrix) occurring at $\Delta = \Delta_1 - \Delta_2 \approx 3.78.$ In all figures, the pink (blue) curve represents the leading edge of the front (trajectory of threshold crossing point) for $j=1$ ($j=2$). }
\label{det_decup}
\end{figure}

\begin{figure}[tb]
\begin{center}
\includegraphics[width=12cm]{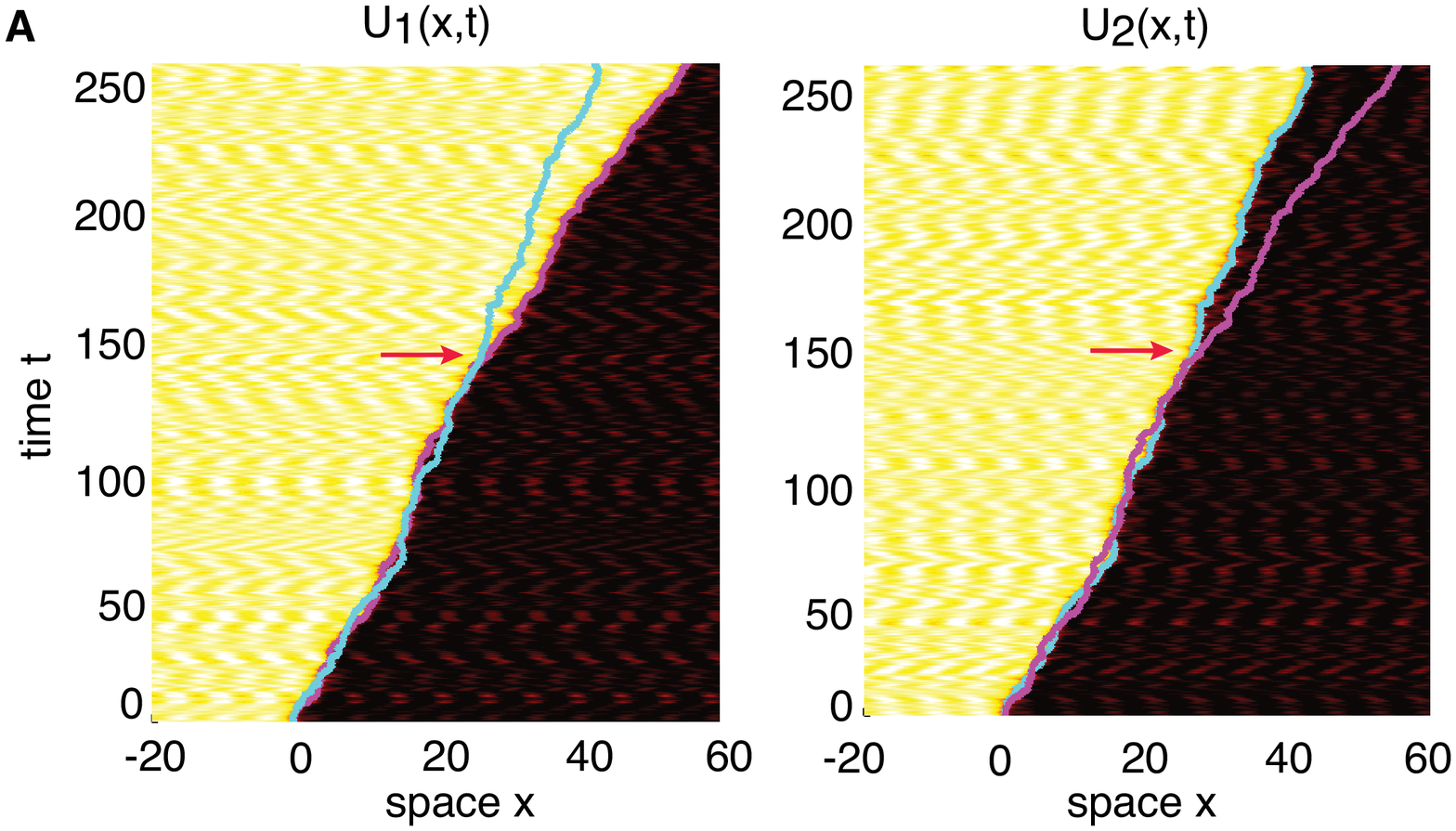} 
\includegraphics[width=8cm]{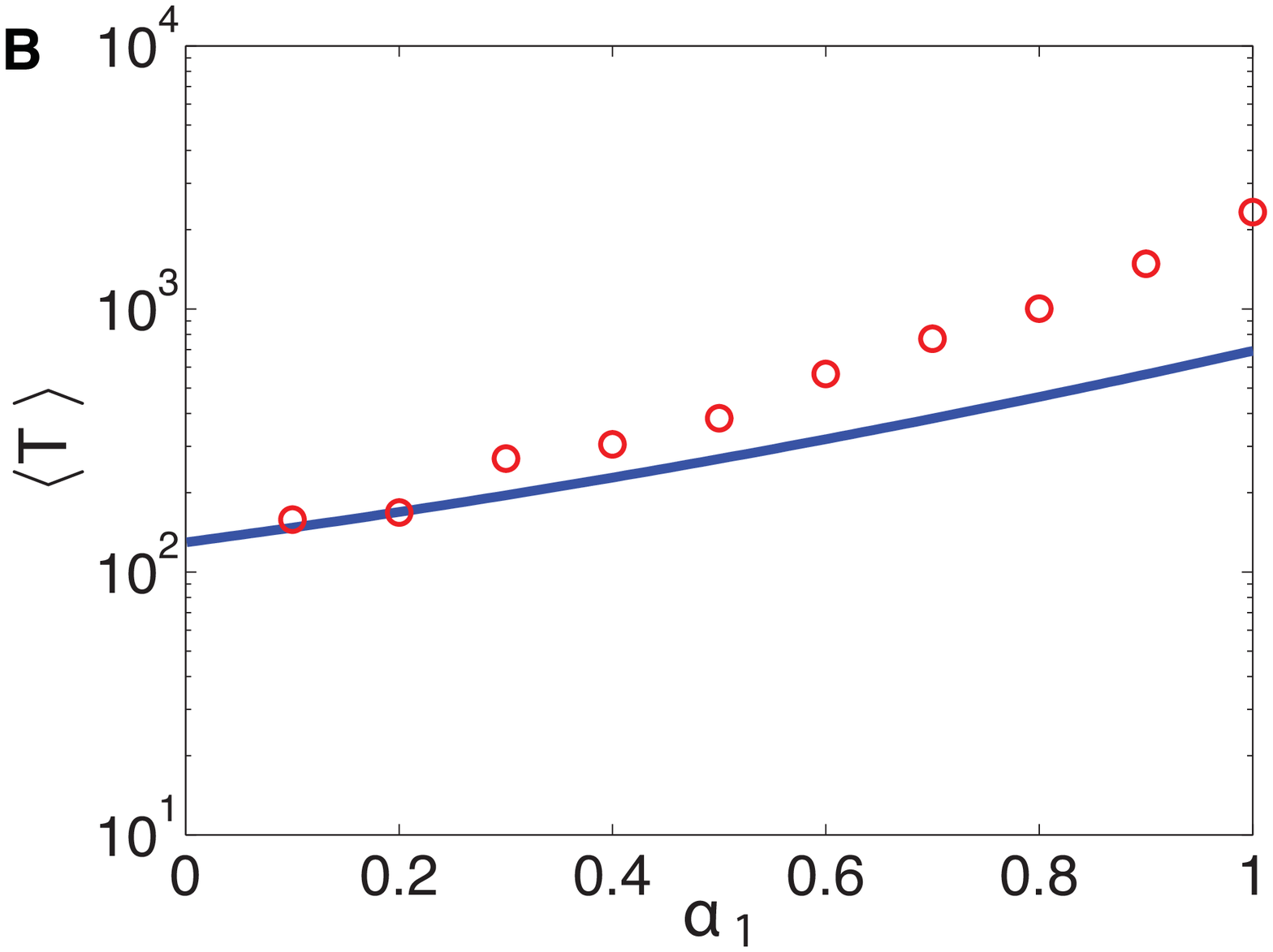}
\end{center}
\caption{\small ({\bf A}) Numerical simulation demonstrating a phase-slip (arrows) for two coupled fronts in the system (\ref{FT1}-\ref{FT2}), using the exponential coupling function (\ref{cupexp}) with $\alpha_1 = \alpha_2 = 0.1$. Overlaid lines represent the evolution of the leading edge of the fronts in time $\Delta_1(t)$ and $\Delta_2(t)$. ({\bf B}) Average time $\tau (0) = \langle T \rangle$ until a phase-slip increases as a function of the interlaminar coupling strength $\alpha_1 = \alpha_2$. Theory (solid line) computed using (\ref{tau0front}) is compared with numerical simulation results (circles) as the coupling $\alpha_1$ is varied.  Threshold $\kappa = 0.4$, noise amplitude $\epsilon = 0.003$. Mean first passage times $\langle T \rangle$ are computed with 1000 samples.}
\label{breakdown_sim}
\end{figure}

Let us focus on the regime where the stable phase-locked state at zero coexists with a pair of unstable states. If we now carry out a linear-noise approximation about the stable state, then the resulting OU process will capture the Gaussian-like fluctuations within the basin of attraction of the phase-locked state on intermediate time-scales. However, on longer time-scales, large fluctuations (rare events) can lead to an escape from the basin of attraction due to $\Delta(t)$ crossing one of the unstable fixed points $\pm \xi_1$. Destabilization of the noise-free system is illustrated in Fig. \ref{det_decup}. Noise-induced escape from the stable state (as in Fig. \ref{breakdown_sim}{\bf A}) cannot be captured using the linear-noise approximation. The full SDE for $\Delta(t)$ is obtained by subtracting equations (\ref{SODE2}a) and (\ref{SODE2}b), which yields
\begin{equation}
\label{gog}
d\Delta(t)=-  G_-(\Delta)dt+\sqrt{\mu}dW(t),
\end{equation}
where we have set $4D=1,\mu= \sqrt{\epsilon}$ and rescaled time according to $t\rightarrow \mu t$. Here
\[\langle dW(t)\rangle =0,\quad \langle dW(t)dW(t')\rangle=\delta(t-t')dt'dt.\]
Let $p(\Delta,t)$ be the probability density for the stochastic process $\Delta(t)$ given some initial condition $\Delta(0)=\Delta_0$. The corresponding Fokker-Planck (FP) equation is given by
\begin{equation}
\label{FP}
\frac{\partial p}{\partial t}= \frac{\partial [G_-(\Delta)p(\Delta,t)]}{\partial \Delta}+\frac{\mu}{2}\frac{\partial^2 p(\Delta,t)}{\partial \Delta^2}\equiv -\frac{\partial J(\Delta,t)}{\partial \Delta},
\end{equation}
where
\[J(\Delta,t)=-\frac{\mu}{2}\frac{\partial p(\Delta,t)}{\partial \Delta}- G_-(\Delta)p(\Delta,t)\]
and $p(\Delta,0)=\delta(\Delta-\Delta_0)$.
Suppose that the deterministic equation $\dot{\Delta}=-G_-(\Delta)$ has a stable fixed point at $\Delta=0$ and a pair of unstable fixed points at $\Delta=\pm \xi_1$. Thus the basin of attraction of the zero state is given by the interval $(-\xi_1,\xi_1)$. For small but finite $\mu$, fluctuations can induce rare transitions out of the basin of attraction due to a metastable trajectory crossing one of the points $\pm \xi_1$. Assume that the stochastic system is initially at $\Delta_0=0$. In order to solve the first passage time problem for escape from the basin of attraction of the zero fixed point, we impose absorbing boundary conditions at $\pm \xi_1$, that is, we set $p(\pm \xi_1,t)=0$.
Let $T(\Delta)$ denote the (stochastic) first passage time for which the system first reaches one of the points $\pm \xi_1$, given that it started at $\Delta\in (-\xi_1,\xi_1)$. The distribution of first passage times is related to the survival probability that the system hasn't yet reached $\pm \xi_1$:
\begin{equation}
S(t)\equiv \int_{ -\xi_1}^{\xi_1}p(\Delta,t)d\Delta .
\end{equation}
That is, $\mbox {Prob}\{t>T\} =S(t)$ and the first passage time density is
\begin{equation}
f(t)=-\frac{dS}{dt}=-\int_{ -\xi_1}^{\xi_1} \frac{\partial p}{\partial t}(\Delta,t)d\Delta .
\end{equation}
Substituting for $\partial p/\partial t$ using the FP equation (\ref{FP}) shows that
\begin{eqnarray}
f(t)&=&\int_{ -\xi_1}^{\xi_1}\frac{\partial J(\Delta,t)}{\partial \Delta}d\Delta= J(\xi_1,t)-J(-\xi_1,t).
\end{eqnarray}
The first passage time density can thus be interpreted as the total probability flux leaving the basin of attraction through the absorbing boundaries.

Using standard analysis, one can show that the mean first passage time (MFPT) $\tau(\Delta)=\langle T\rangle$ satisfies the backwards equation \cite{Gardiner09,vanKampen92}
\begin{equation}
- G_-(\Delta)\frac{d\tau}{d \Delta}+\frac{\mu}{2}\frac{d^2 \tau}{d\Delta^2}=-1,
\end{equation}
with the boundary conditions
$\tau(\pm\xi_1)=0$. Solving this equation yields
\begin{align}
\tau(\Delta)=\tau_1(\Delta)-\tau_2(\Delta)
\end{align}
with
\begin{align}
\tau_1(\Delta)=\frac{2}{\mu}\left (\int_{-\xi_1}^{\xi_1}\frac{dy}{\psi(y)}\right )^{-1}\left (\int_{-\xi_1}^{\Delta}\frac{dy}{\psi(y)}\right )\left [\int_{\Delta}^{\xi_1}\frac{dy'}{\psi(y')}\int_{-\xi_1}^{y'}\psi(z)dz\right ],
\end{align}
\begin{align}
\tau_2(\Delta)=\frac{2}{\mu}\left (\int_{-\xi_1}^{\xi_1}\frac{dy}{\psi(y)}\right )^{-1}\left (\int_{\Delta}^{\xi_1}\frac{dy}{\psi(y)}\right )\left [\int_{-\xi_1}^{\Delta}\frac{dy'}{\psi(y')}\int_{-\xi_1}^{y'}\psi(z)dz\right ],
\end{align}
and
\begin{align}
\psi(\Delta)=\exp\left [-\frac{2}{\mu}\int_{0}^{\Delta} G_-(y)dy\right ].
\end{align}
We can simplify the analysis considerably by exploiting the symmetry of the given problem, namely, that $G_-(\Delta)=-G_-(-\Delta)$ and hence $\psi(\Delta)=\psi(-\Delta)$.
It is then straightforward to show that
\begin{align}
\tau(0)=\frac{2}{\mu}\int_{0}^{\xi_1}\frac{dy'}{\psi(y')}\int_{0}^{y'}\psi(z)dz.  \label{tau0front}
\end{align}
This is identical to the formula for the MFPT for escape from the interval $[0,\xi_1)$ starting at $\Delta=0$ with a reflecting boundary at $\Delta=0$ and an absorbing boundary at $\Delta=\xi_1$. We compare our theory for the mean first passage time $\langle T \rangle = \tau (0)$ to numerical simulations for symmetric lateral inhibitory connectivity (\ref{dog}) and cosine noise correlations so $D$ satisfies (\ref{Decos}), showing in Fig. \ref{breakdown_sim} that the marginal rare event statistic is captured for low enough values of coupling strength $\alpha_1$. However, as expected, our theory breaks down as the coupling strength $\alpha_1$ is increased. Our perturbation analysis is built upon the assumption of weak coupling between layers, so the reduced system will be a poorer characterization of the full integrodifferential system's dynamics for stronger coupling. For example, our reduced system does not account for the reshaping of the front profile by the interlaminar coupling, which likely impacts the waiting time until front decoupling. Fronts' shapes may be altered in ways such that there are multiple crossings of the the firing rate threshold $\kappa$, and capturing this feature would require a more detailed reduced system.

Using steepest descents, one can now derive the classical Arrhenius formula \cite{Gardiner09}
\begin{equation}
 \tau(0)\sim \frac{2\pi }{\sqrt{|U''(\xi_1)|U''(0)}}\e^{2 [U(\xi_1)-U(0)]/\mu},
 \label{arrh}
 \end{equation}
where
 \begin{equation}
 U(\Delta)=\int_0^{\Delta}G_-(y)dy.
 \end{equation}
 
The above analysis provides another example where knowledge of the nonlinear nature of the phase-shift dynamics is crucial for determining the effects of noise. In \S 2 and \S 3, the nonlinearity determined the unique phase-locked state $\xi_0$ about which we could carry out a linear-noise approximation. Here, the explicit form of the nonlinearity is required in order to determine the rate of escape from a (meta)stable phase-locked state that coexists with a pair of unstable states. Note that, as expected,  the mean escape time is exponentially large in the weak noise limit $\mu \rightarrow 0$, since $U(\xi_1)>U(0)$.

\section{Bumps in weakly-coupled stochastic neural fields}
\label{cupbumps}

Kilpatrick has previously shown that interlaminar coupling can reduce the long term diffusion of bumps in stochastic neural fields \cite{Kilpatrick13a,Kilpatrick14a}. Effective equations for the stochastic motion of bumps derived in this work tended to be linear SDEs, such as OU processes. However, in \cite{Kilpatrick13} it was shown how the impact of spatial heterogeneities can be accounted for by a nonlinear SDE that incorporates the effective potential bestowed by the heterogeneity. Here, we extend this previous work as well as our analysis of coupled fronts by considering nonlinear contributions to the effective dynamics of bumps coupled in laminar neural fields. We focus on a pair of identical neural fields on the ring $x \in (- \pi, \pi]$, labeled $j=1,2$ and mutually coupled by interlaminar weight distributions $J_1(x)$ and $J_2(x)$:
\begin{subequations}
\begin{align}
\tau d U_1(x,t) =& \left[ - U_1(x,t) + \int_{- \pi}^{\pi} w(x-y) F(U_1(y,t)) dy \right] dt \label{BT1} \\
& \qquad + \epsilon^{1/2} \int_{- \pi}^{\pi} J_1(x-y) F(U_2(y,t)) dy dt + \epsilon^{1/2} \tau^{1/2} dW_1(x,t), \nonumber
\end{align}
and
\begin{align}
\tau d U_2(x,t) = & \left[ - U_2(x,t) + \int_{- \pi}^{\pi} w(x-y) F(U_2(y,t)) dy \right] dt  \label{BT2} \\
& \qquad + \epsilon^{1/2} \int_{- \pi}^{\pi} J_2(x-y) F(U_1(y,t)) dy dt + \epsilon^{1/2} \tau^{1/2} dW_2(x,t).
\end{align}
\end{subequations}
The spatially extended Wiener processes $W_1(x,t)$ and $W_2(x,t)$ are defined on $x \in ( -\pi ,\pi]$ with
\begin{align}
\langle dW_j(x,t) \rangle = 0, \hspace{5mm} \langle dW_i (x,t) dW_j(x',t) \rangle = 2 C (x-x') \delta_{i,j}\delta ( t- t') dt dt'.
\end{align}
Coupling within a layer $w(x-y)$ is assumed to be an even function of lateral inhibitory type \cite{Amari77,Coombes04,Kilpatrick13}. For comparison with numerical simulations, we assume intralaminar coupling
\begin{align}
w(x) = \cos (x) \label{wcos}
\end{align}
and interlaminar coupling $J_j(x)$ may be asymmetric
\begin{align}
J_j(x) = \alpha_j \cos (x),  \label{intercos}
\end{align}
and we will allow $\alpha_j$ to be negative or positive. We will also set the timescale $\tau = 1$. 

In the absence of interlaminar coupling and noise ($\epsilon = 0$), it can be shown that each neural field independently supports a stationary pulse solution (bump), which satisfies the equation
\begin{align}
U_0(x) = \int_{- \pi}^{\pi} w(x-y) F(U_0(y)) dy  \label{bsol}
\end{align}
(see \cite{Kilpatrick13} for details). Each neural field is translationally symmetric, so each bump $U_0(x)$ is centered at an arbitrary location in the neural field. The nonlinear effects of noise and weak coupling upon the position of the bumps can then be analyzed along identical lines to that of traveling fronts in \S 3, after setting $c=0$ and taking $x\in [-\pi,\pi]$ rather than $x\in \Rset$. In particular, writing
\begin{align}
U_j(x,t) = U_0(x- \Delta_j(t)) + \epsilon^{1/2} \Phi_j(x- \Delta_j(t),t), \hspace{5mm} j=1,2,  \label{bansatz}
\end{align}
we find that the displacements $\Delta_j(t)$ satisfy the Langevin equation {(\ref{SODE2}) with ${\mathcal V}(\x)$ now the null vector of
the non-self-adjoint linear operator
\begin{align}
\widehat{L} A(x)= -A(x) +\int_{-\pi}^{\pi} w(x-x')F'(U_0(x'))A(x')dx'
\end{align}
for any function $A(x) \in L^2( [-\pi,\pi])$ and $U_0$ a bump solution. (The integrals in equations (\ref{lamG}), (\ref{gogo}) and (\ref{D2}) are now taken with respect to the circle rather than the real line.) Continuing along identical lines to \S 3, we  can show that the phase difference $\Delta(t)=\Delta_1(t)-\Delta_2(t)$ evolves according to equation (\ref{detD}). Hence, a stable phase locked state $\Delta(t)=x_0$ will satisfy $G_-(x_0) = 0$ and $G'(x_0)>0$. Finally, carrying out a linear noise approximation establishes that fluctuations about the phase-locked state are given by an OU process, whereas fluctuations in the center-of-mass are given by Brownian diffusion in the large-time limit, as was previously found in \cite{Kilpatrick13a}. Moreover, in the limit $t\rightarrow \infty$,
\begin{align}
\langle \Delta (t)^2 \rangle - \langle \Delta (t) \rangle^2 = \frac{2 \sqrt{\epsilon} D}{G_-'(x_0)} , \label{bpdiff0}\end{align}
with
\begin{align}
D = \frac{ \int_{- \pi}^{\pi} \int_{- \pi}^{\pi} {\mc V}(x) C (x - x') {\mc V}(x') d x' d x}{\left[ \int_{- \pi}^{\pi} {\mc V}(x) U_0'(x) d x \right]^2}.  \label{bumpdiff}
\end{align}
and $G_-(\Delta)=G_1(\Delta)-G_2(-\Delta)$ for
\begin{align}
G_j ( \Delta ) = \frac{\int_{- \pi}^{\pi} {\mc V}(x) \left[ \int_{- \pi}^{\pi} J_j(x - x') F(U_0(x' + \Delta )) \d x' \right] d x}{\int_{- \pi}^{\pi} {\mc V}(x) U_0'(x) d x}.  \label{Gjbump}
\end{align}

\subsection{Explicit results for Heaviside rate function} The uncoupled deterministic neural field equations ($\epsilon = 0$) are given by
\begin{align}
\frac{\pd u_j(x,t)}{\pd t} = - u_j(x,t) + \int_{- \pi}^{\pi} w(x-y) H(u_j(y,t) - \kappa) dy, \hspace{5mm} j=1,2.
\end{align}
Thus, for a cosine weight function (\ref{wcos}), bump solutions satisfy (\ref{bsol}), so fixing their peak to be at $x=0$, we have \cite{Kilpatrick13}
\begin{align}
U_0(x) = A \cos x, \hspace{5mm} U_0'(x) = -A \sin x, \hspace{5mm} A= \sqrt{1 + \kappa^2} - \sqrt{1 - \kappa^2},
\end{align}
and
\begin{align}
U_0( \pm a) = \kappa, \hspace{5mm} a = \frac{\pi}{4} \pm \left( \frac{\pi}{4} - \frac{1}{2} \sin^{-1} \kappa \right),
\end{align}
where the wider bump (larger $a$) will be stable \cite{Kilpatrick13}. The null-space ${\mc V}(x)$ of the adjoint operator $\widehat{L}^*$ satisfies
\begin{align}
{\mc V}(x) = \frac{\delta(x+a)}{|U'(a)|} \int_{- \pi}^{\pi} \cos (a+y) {\mc V}(y) dy + \frac{\delta(x-a)}{|U'(a)|} \int_{- \pi}^{\pi} \cos (a-y) {\mc V}(y) dy,
\end{align}
which has solution ${\mc V}(x) = \delta (x+a) - \delta (x-a)$. Fo concreteness, we take $C(x) = C_0 \cos (x)$, where $C_0$ has units of length. It follows from equation (\ref{bumpdiff}) that
\begin{align}
D&= C_0 \frac{\left( \int_{- \pi}^{\pi} {\mc V} (x) \sin (x) dx \right)^2 + \left( \int_{- \pi}^{\pi} {\mc V} (x) \cos (x) dx \right)^2  }{ \left[ U_0'(a) - U_0'(-a) \right]^2} = \frac{C_0}{2 + 2 \sqrt{1- \kappa^2}}.  \label{Dbcos}
\end{align}
Moreover, noting that
\begin{align}
\int_{- \pi}^{\pi} {\mc V}(x) U_0'(x) dx = U_0'(-a) - U_0'(a) = 2A \sin a
\end{align}
and taking cosine interlaminar connectivity (\ref{intercos}), we have
$G_j( \Delta ) = \alpha_j \sin \Delta$,
so that
\begin{align}
G_{\pm}(\Delta) = ( \alpha_1 \mp \alpha_2 ) \sin \Delta.   \label{Gpmsine}
\end{align}
A related result was derived in \cite{Kilpatrick13}, demonstrating that weak nonlinear spatial heterogeneities can be inherited by the underlying dynamics of stochastically moving bumps in single layer neural fields. Hence, w can immediately identify the two fixed points on $\Delta \in ( - \pi , \pi]$, where $G_-( \Delta) \equiv 0$ at $\bar{\Delta}=0, \pi$. Their stability is easily computed
\begin{align}
\lambda_0 & \equiv -G_-'(0) = - ( \alpha_1 + \alpha_2), \\
\lambda_{\pi} & \equiv -G_-'(\pi) = ( \alpha_1 + \alpha_2),
\end{align}
so if $\alpha_1 + \alpha_2 >0$, $x_0 = 0$ ($x_1 = \pi$) is stable (unstable), and if $\alpha_1 + \alpha_2 <0$, $x_1 = 0$ ($x_0 = \pi$) is unstable (stable). Thus, for locally inhibitory connectivity, the bumps' positions can be driven apart ($x_0 = \pi$) so they are anti-phase relative to one another. Either way, we can compute the stationary variance about the phase-locked state
\begin{align}
\langle \Delta (t)^2 \rangle - \langle \Delta (t) \rangle^2 = \frac{2 \sqrt{\epsilon} D}{G_-'(x_0)} = \frac{C_0 \sqrt{\epsilon}}{|\alpha_1 + \alpha_2| (1 + \sqrt{1- \kappa^2})}. \label{bpdiff}\end{align}
Thus, we recover the result from \cite{Kilpatrick13a}, showing the strength of interlaminar coupling reduces variance in stochastic bump motion. We demonstrate in Fig. \ref{bvarplot} that the variance in the phase difference is well approximated by (\ref{bpdiff}) as compared with numerical simulations.

\begin{figure}[tb]
\begin{center} \includegraphics[width=6.3cm]{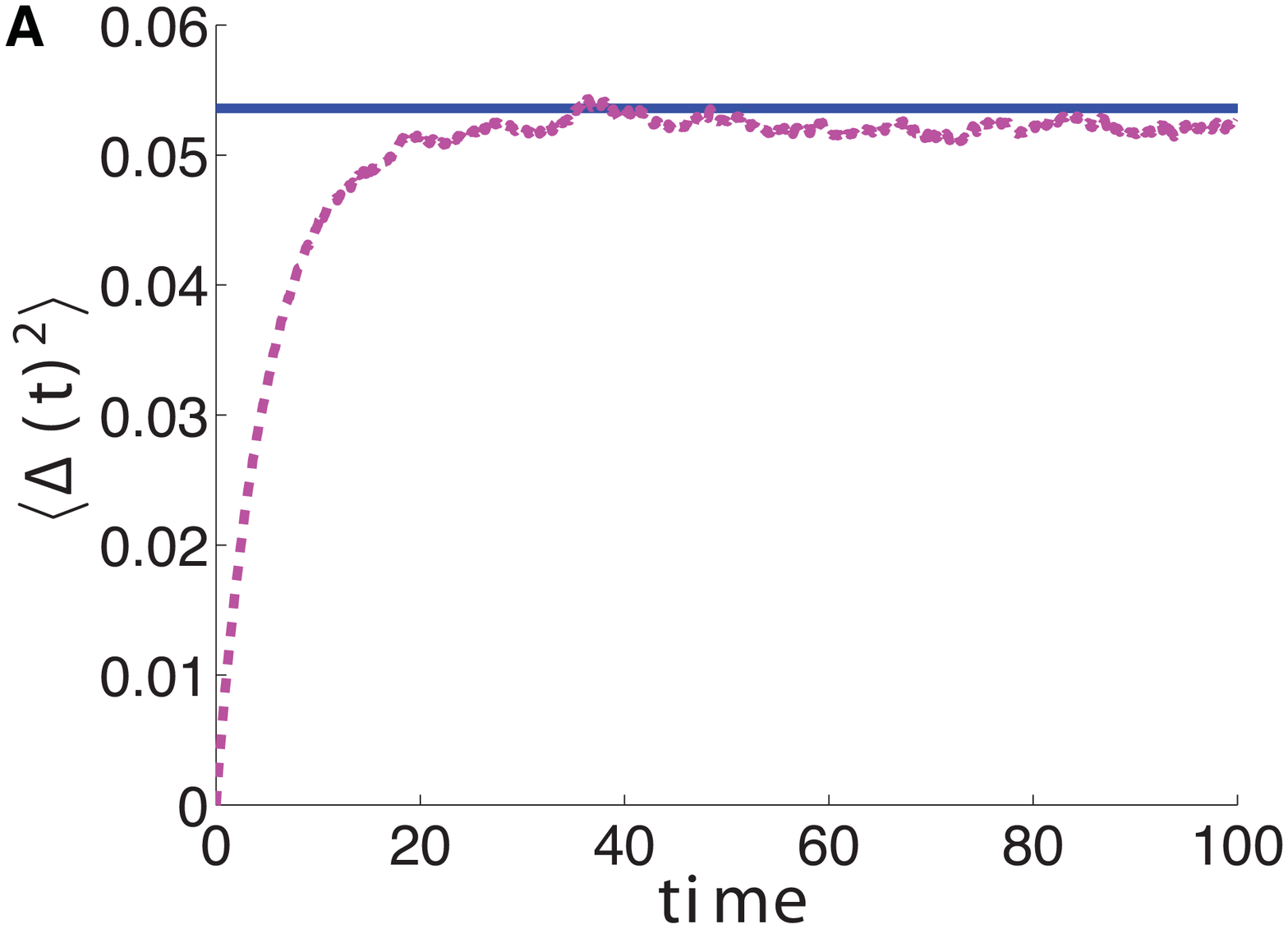} \includegraphics[width=6.3cm]{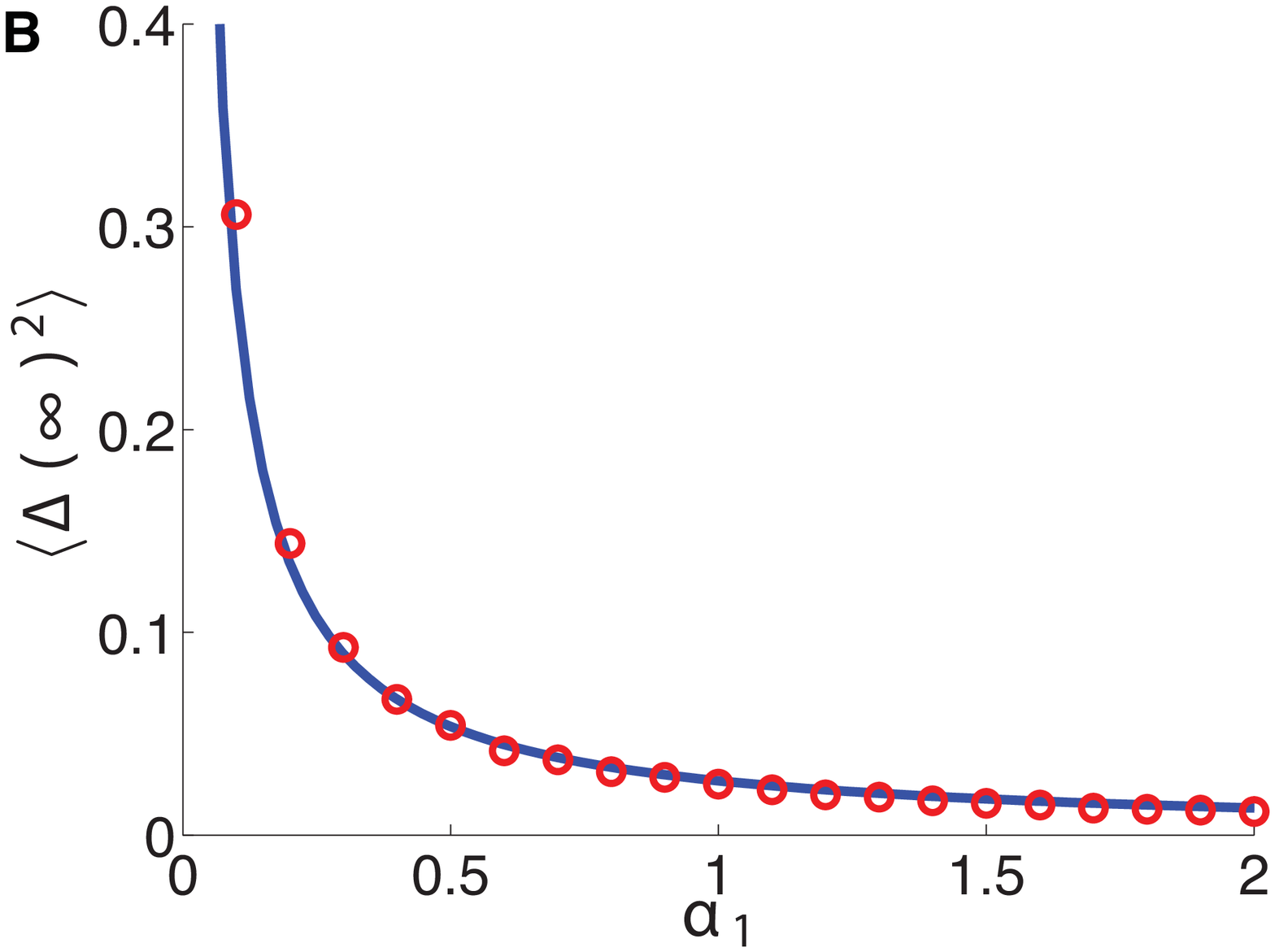} \end{center}
\caption{\small {\bf A} The variance in the phase difference $\langle \Delta^2 \rangle$ (note the mean $\langle \Delta \rangle \equiv 0$) saturates in the limit $t \to \infty$. Variance is computed as a function of time using numerical simulations (dashed line) and asymptotic theory (solid line) predicts the saturation value (\ref{bpdiff}). Threshold $\kappa = 0.5$; interlaminar connectivity strength $\alpha_1 = \alpha_2 = 1$; noise amplitude $\epsilon = 0.01$. {\bf B} The stationary variance decreases as a function of interlaminar connectivity strength $\alpha_1 = \alpha_2$ in numerical simulations (circles) and theory (solid line). Other parameters are as in {\bf A}. Variances are computed using 5000 realizations each.}
\label{bvarplot}
\end{figure}

\subsection{Noise-induced phase slips} As in the case of fronts, noise not only causes the center-of-mass of both bumps to wander diffusively, it can also lead to rare events, where the bumps temporarily become uncoupled. Typically, noise will perturb the phase-difference $\Delta (t)$, while the local dynamics of the stable fixed point will pull the phase difference back to $x_0$. However, on longer time-scales, large fluctuations can lead to $\Delta (t)$ crossing the separatrix given by the unstable fixed point $x_1$. These events cannot be captured by a linear noise approximation. Note, a similar observation was made by Kilpatrick and Ermentrout in \cite{Kilpatrick13} in a single layer network with periodic spatial heterogeneity in the weight function.

The nonlinear SDE describing the stochastic motion of the phase-difference $\Delta (t)$ is given by equation (\ref{gog}).
Let $p( \Delta, t)$ be the probability density of the stochastic process $\Delta (t)$ given initial condition $\Delta (0) = \Delta_0$. The corresponding Fokker-Planck (FP) equation takes the form
\begin{equation}
\label{FPbump}
\frac{\pd p}{\pd t}= \frac{\pd [G_-(\Delta)p(\Delta,t)]}{\pd \Delta}+\frac{\mu}{2}\frac{\pd^2 p(\Delta,t)}{\pd \Delta^2}\equiv -\frac{\partial J(\Delta,t)}{\partial \Delta},
\end{equation}
where
\[J(\Delta,t)=-\frac{\mu}{2}\frac{\partial p(\Delta,t)}{\partial \Delta}- G_-(\Delta)p(\Delta,t)\]
and $p(\Delta,0)=\delta(\Delta-\Delta_0)$.
As in the case of fronts, we suppose that the deterministic equation $\dot{\Delta}=-G_-(\Delta)$ has a stable fixed point at $\Delta=0$ and a pair of unstable fixed points at $\Delta=\pm x_1$. Thus the basin of attraction of the zero state is given by the interval $(-x_1,x_1)$. For small but finite $\mu$, fluctuations can induce rare transitions where trajectories cross through one of the unstable separatrices $\pm x_1$. To solve the first passage time problem for these phase-slips, starting with $\Delta (0) = 0$, we impose absorbing boundary conditions at $\pm x_1$, that is, we set $p(\pm x_1,t)=0$.
Let $T(\Delta)$ denote the (stochastic) first passage time for which the system first reaches one of the points $\pm x_1$, given that it started at $\Delta\in (-x_1,x_1)$. As in the case of fronts, one can show that the mean first passage time (MFPT) $\tau(0)=\langle T (0) \rangle$ is given by the formula
\begin{align}
\tau(0)=\frac{2}{\mu}\int_{0}^{x_1}\frac{dy'}{\psi(y')}\int_{0}^{y'}\psi(z)dz,
\end{align}
where
\begin{align}
\psi(\Delta)=\exp\left [-\frac{2}{\mu}\int_{0}^{\Delta} G_-(y)dy\right ].
\end{align}

\begin{figure}[tb]
\begin{center} \includegraphics[width=12cm]{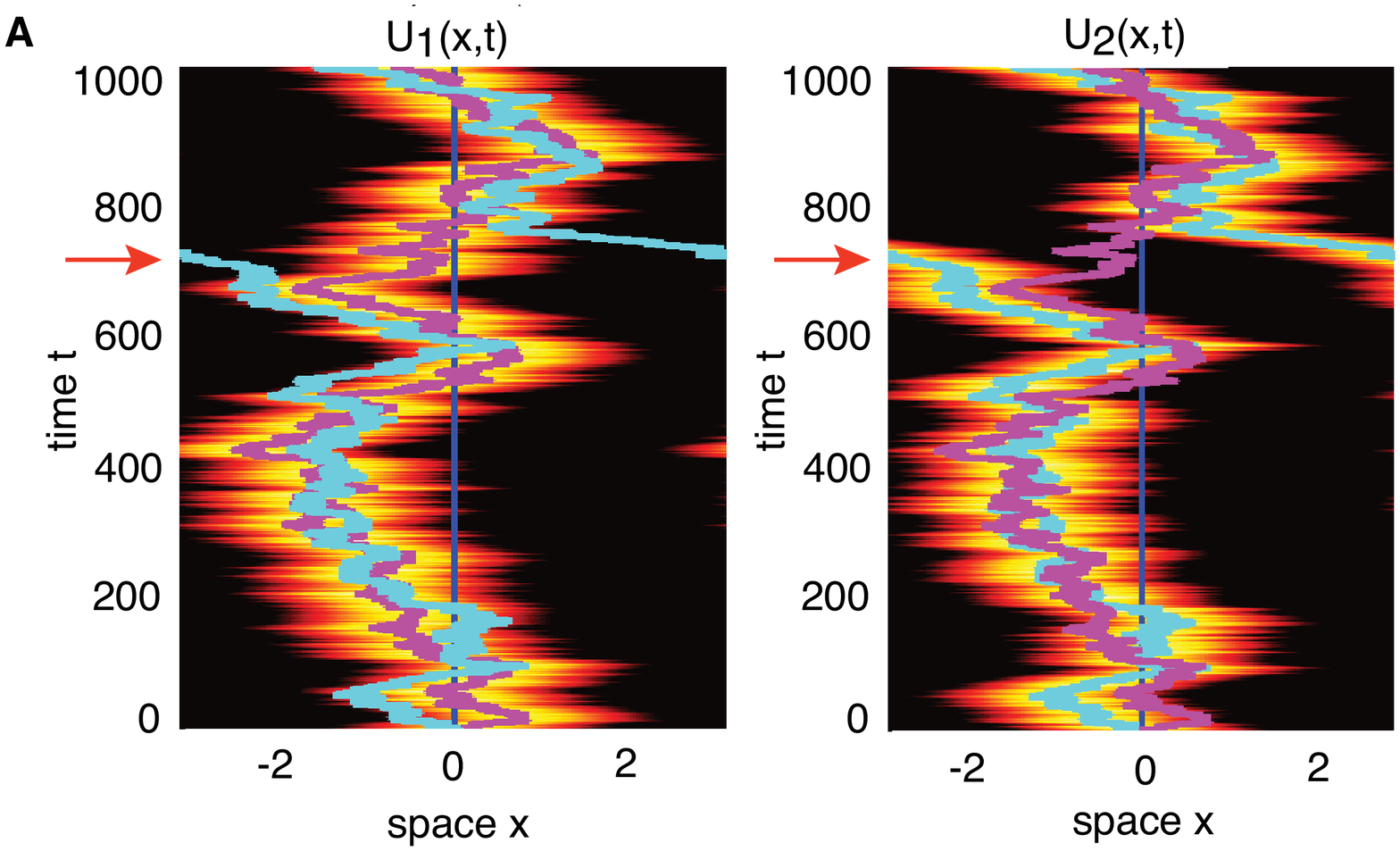}  \includegraphics[width=8cm]{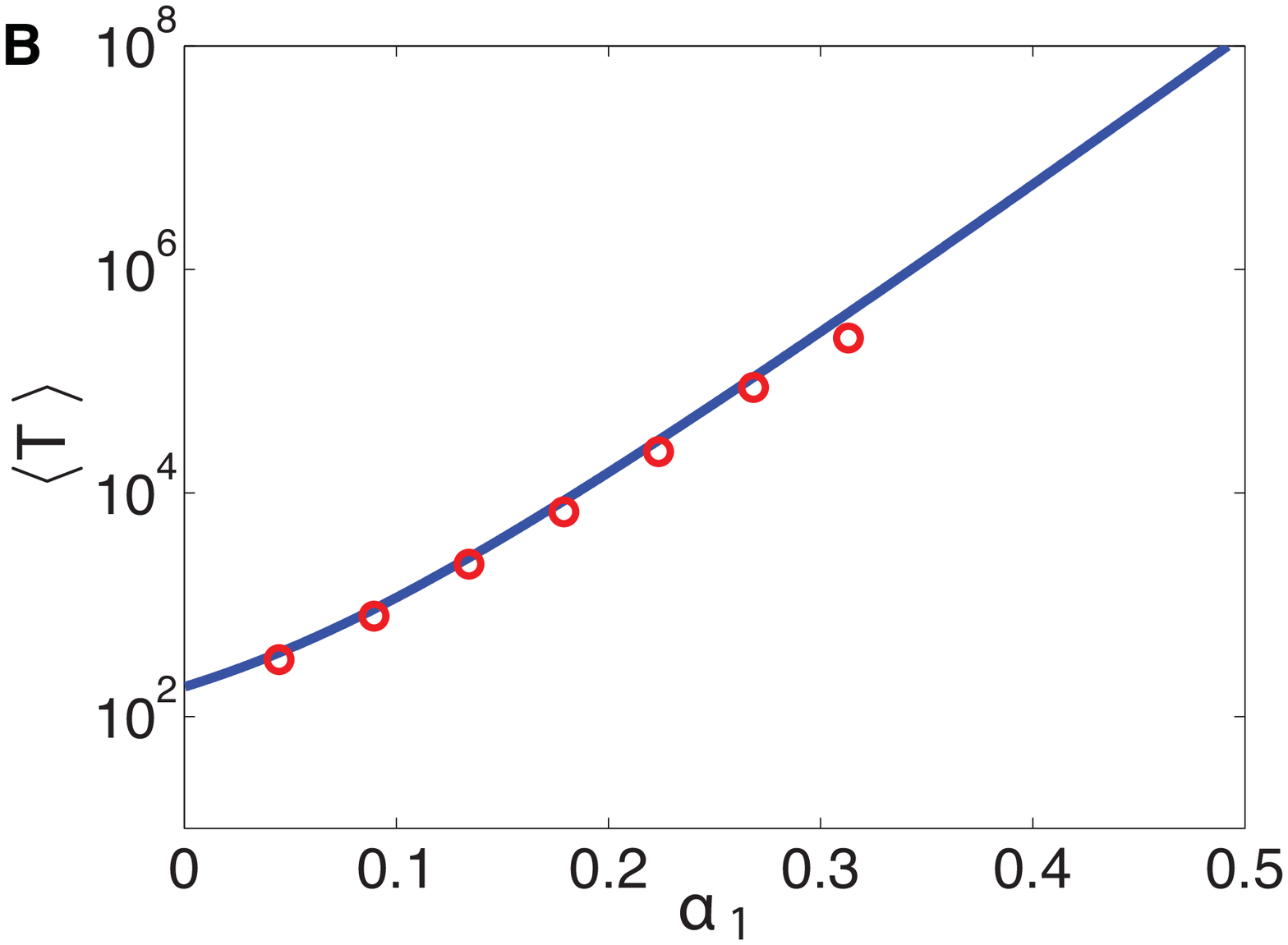} \end{center}
\caption{\small ({\bf A}) Numerical simulation demonstrating a phase-slip (arrows) for two coupled bumps in the system (\ref{BT1}-\ref{BT2}), using the cosine coupling function (\ref{intercos}) with $\alpha_1 = \alpha_2 = 0.1$. Overlaid lines represent the evolution of centers-of-mass of the bumps in time $\Delta_1(t)$ and $\Delta_2(t)$. ({\bf B}) Average time $\tau (0) = \langle T \rangle$ until a phase slip increases as a function of the interlaminar coupling strength $\alpha_1 = \alpha_2$. Theory (solid line) computed using (\ref{coscomptau0}) matches numerical simulation results (circles) well.  Threshold $\kappa = 0.5$, noise amplitude $\epsilon = 0.05$. Mean first passage times $\langle T \rangle$ are computed with 1000 samples.}
\label{sliptime}
\end{figure}

We can obtain explicit results in the case of a Heaviside rate function, local cosine weight function (\ref{wcos}), and cosine interlaminar connectivity (\ref{intercos}). To begin, we employ equation (\ref{Gpmsine}) to note that
\begin{align}
\psi(\Delta)=\exp\left [-\frac{2 (\alpha_1 + \alpha_2)}{\mu}\int_{0}^{\Delta} \sin (y)dy\right ] = \exp\left [- \bar{\alpha} (1 - \cos \Delta) \right ],
\end{align}
where $\bar{\alpha} = 2 (\alpha_1 + \alpha_2)/\mu$. Thus,
\begin{align}
\tau (0) = \frac{2}{\mu} \int_0^{\pi} \int_0^x \e^{\displaystyle \bar{\alpha} (\cos y - \cos x)} dy dx,  \label{tslipbump}
\end{align}
which is straightforward to integrate using numerical quadrature. We compare the analytically derived approximation to the mean time between phase slips (\ref{tslipbump}) to results computed numerically simulating the full system of stochastic integrodifferential equations (\ref{BT1}-\ref{BT2}) in Fig. \ref{sliptime}. The theory we compare assumes cosine noise correlations ($C(x) = C_0\cos (x)$), so that $D$ is given by (\ref{Dbcos}). Assuming symmetric coupling $\alpha_1=\alpha_2 = \alpha$ and inverting the rescaling $t \to \mu^{-1} t$, we have
\begin{align}
\tau (0) = \frac{1}{2 \epsilon D}  \int_0^{\pi} \int_0^x \e^{\displaystyle \frac{\alpha}{\sqrt{\epsilon} D} (\cos y - \cos x)} dy dx.  \label{coscomptau0}
\end{align}

\section{Discussion}

In this paper we have explored the impact of stimuli and coupling on the stochastic motion of patterns in neural field equations. Our main advance is to demonstrate nonlinear contributions to the effective stochastic equations for the position of wandering patterns. In order to achieve this, we have assumed that the stimuli or the coupling between multiple layers of a neural field are weak, namely having comparable amplitude to the spatiotemporal noise term that forces the system. Then, utilizing a small noise and small coupling expansion, we have derived nonlinear Langevin equations whose effective potentials are shaped by the spatial profile of coupling. This allows us to approximate the statistics of rare events, such as large deviations whereby waves become decoupled from one another, analogous to the well-hopping of bumps observed in spatially heterogeneous neural fields \cite{Kilpatrick13}. Such stochastic dynamics would not be captured by a linear system of SDEs.

As with all perturbation methods, our results are dependent on the particular choice of scaling with respect to $\epsilon$. We have chosen the scaling of the external inputs, noise amplitudes and interlaminar coupling so that all the terms contribute to the amplitude equation obtained by applying the Fredholm alternative. If one changed the scaling then one or more of these contributions would disappear at this level of perturbation theory. Unfortunately, not enough is known biologically to determine the ``correct'' scaling. However, there are biological reasons for treating these various contributions as weak (see the review \cite{Bressloff12a}). First, neural systems tend to adapt to constant stimuli and amplify weak fluctuations about a constant background via recurrent connections (weak external inputs). Second,  there are long-range connections within cortex that play a modulatory role in neuronal firing patterns (weak interlaminar connections). Third, although individual cortical neurons tend to be noisy, at the population level fluctuations tend to be relatively weak (weak noise).

In \S3--\S5 we considered a pair of one-dimensional (1D) neural fields that are homogeneous when $\epsilon =0$ (no interlaminar coupling nor external noise). This means that each unperturbed neural field supports a 1D pattern (front, bump) with a marginally stable degree of freedom (also known as a Goldstone mode), which reflects the underlying translation symmetry. The incorporation of weak interlaminar coupling  and noise breaks the translation symmetry of each neural field, resulting in a pair of nonlinear Langevin equation for the positions of the two patterns. Carrying out a linear noise approximation of the dynamics in the neighborhood of a phase-locked state shows that the center-of-mass (CoM) of the two patterns exhibits Brownian dynamics, whereas the separation of the two patterns satisfies an OU process. A natural generalization of our analysis is to consider $N$ neural fields. There are now $N$ Goldstone modes when $\epsilon =0$, arising from the fact that each neural field is equivariant with respect to local uniform shifts. Introduction of weak pair-wise interlaminar coupling and noise would result in $N$ coupled Langevin equations, specifying the individual positions of the $N$ coupled patterns. Again, under a linear noise approximation, we expect the CoM of the $N$ patterns to exhibit Brownian dynamics, whereas the $N-1$ remaining degrees of freedom satisfy an OU process as shown in \cite{Kilpatrick13a}. The existence of Brownian dynamics is a consequence of the fact that in the presence of interlaminar coupling but no noise, the $N$ Goldstone modes reduce to a single Goldstone mode representing translation symmetry with respect to a global shift of all the fields.  Addition of a weak external input to one of the neural fields would break this remaining symmetry, and the CoM could phase-lock to the stimulus. Under a linear noise approximation, the CoM would now satisfy an OU process. 

Another obvious extension of our theory would be to consider two-dimensional (2D) neural fields, reflecting the laminar-like structure of cerebral cortex \cite{Lund03}. For example, one could extend the theory of radially symmetric bumps developed in Ref. \cite{Folias04} to the case of multi-layer neural fields. In this case each neural field has two marginally stable modes when $\epsilon =0$, corresponding to translation symmetry in the plane. One could also consider  the stochastic motion of more intricate spatiotemporal patterns such as spiral waves, which occur when some form of slow adaptation is included \cite{Laing05}. In particular, the methods developed here could be employed to study how noise shapes the angular velocity and tip location of spiral waves coupled to external stimuli or other patterns.
Finally, our approach might be utilized to study stochastic switching in neural field models of binocular rivalry \cite{Kilpatrick10,Bressloff12b}. Since switching is brought about by a combination of adaptation as well as fluctuations, multiple timescale methods could be combined with our small noise approximation to calculate the distribution of switching times \cite{Berglund06}. More generally, it would be interesting to develop a large deviation theory for phase-locked patterns in stochastic neural fields, see for example Kuehn and Riedler \cite{Kuehn14}.

\end{document}